%% file: main.tex
\documentclass{article}
\usepackage{spconf,amsmath,graphicx}
\usepackage{url,hyperref}
\usepackage{makecell}
\usepackage{diagbox}
\usepackage{booktabs}
\usepackage{adjustbox}
\usepackage{multirow}
\usepackage{cite}
\usepackage{xcolor}
\usepackage{pgfplots}
\usepackage{tikz}
\usepackage{caption}
\usepackage{subcaption}

\pgfplotsset{width=\linewidth*1.04,compat=1.17} 

\title{Fast Entropy-Based Methods of Word-Level Confidence Estimation for End-To-End Automatic Speech Recognition}
\name{\begin{tabular}{c} Aleksandr Laptev$^{1,2}$, Boris Ginsburg$^1$
\end{tabular}
}
\address{
  $^1$NVIDIA, USA\\
  \vspace{-1pt} 
  $^2$ITMO University, St. Petersburg, Russia}

\copyrightnotice{978-1-6654-7189-3/22/\$31.00~\copyright2023 IEEE}
\toappear{To appear in {\it Proc.\ SLT 2022, Jan 09-12, 2023, Doha, Qatar}}
\begin{document}
%
\maketitle
\begin{abstract}
This paper presents a class of new fast non-trainable entropy-based confidence estimation methods for automatic speech recognition. We show how per-frame entropy values can be normalized and aggregated to obtain a confidence measure per unit and per word for Connectionist Temporal Classification (CTC) and Recurrent Neural Network Transducer (RNN-T) models. Proposed methods have similar computational complexity to the traditional method based on the maximum per-frame probability, but they are more adjustable, have a wider effective threshold range, and better push apart the confidence distributions of correct and incorrect words. We evaluate the proposed confidence measures on LibriSpeech test sets, and show that they are up to 2 and 4 times better than confidence estimation based on the maximum per-frame probability at detecting incorrect words for Conformer-CTC and Conformer-RNN-T models, respectively.
\end{abstract}
\begin{keywords}
\vspace{-1pt} 
ASR, confidence, entropy, CTC, RNN-T
\end{keywords}

\section{Introduction}
\label{sec:intro}
\vspace{-3pt} 

Many real-world automatic speech recognition (ASR) systems require to have not only the best possible transcript, but also confidence scores for each recognized unit, which provide an estimate of how likely the prediction is to be correct. ASR confidence scores are used in semi-supervised and active learning \cite{vesely2013asru,riccardi2005,yu2010csl,drugman2016}, 
speaker adaptation \cite{uebel2001speaker}, information retrieval \cite{zbib2019aS},  speech translation \cite{saleem2004using,besacier-etal-2014-word} etc.
The confidence measures for hybrid WFST-based ASR uses word lattices, $N$-best lists, and external trainable classifiers \cite{wessel2001, Jiang2005ConfidenceMF, Yu2011CalibrationOC}.

This work is motivated by the need for fast, simple, robust, and adjustable confidence estimation for the end-to-end Connectionist Temporal Classification (CTC) \cite{graves_connectionist_2006} and Recurrent Neural Network Transducer (RNN-T) \cite{graves2012transducer} ASR models. 
End-to-end neural ASR systems predict the softmax probability for each unit of the output vocabulary. The probability of the most-likely unit (the maximum probability) is a natural way of estimating confidence \cite{hendrycks2016iclr,park20d_interspeech}. 
The effectiveness of this approach is limited by the so-called \textit{prediction overconfidence}, when the probability distribution is skewed towards the best hypothesis \cite{nguyen2015}. We consider a model to be overconfident when its median probability of incorrect predictions is above 0.9. This situation is typical for overtrained CTC and RNN-T models. To mitigate the overconfidence issue, one can use temperature scaling, dropout, ensemble of ASR models etc  \cite{vyas2019icassp,malinin2021uncertainty, Oneata2021}. An alternative approach is based on dedicated neural confidence models ~\cite{li2020confidence,jeon2020,woodward2020confidence,qiu2021learning,li2021residual,qui2021multi,wang2021word}.



Another challenge in developing confidence estimation methods for end-to-end ASR systems is the granularity of predictions. Speech applications generally require word-level confidence, while end-to-end ASR provides unit-level output. Word-level confidence can either be computed directly as posteriors of a decoding lattice \cite{evermann2000posterior} or they can be aggregated from unit-level scores. The former requires significant computational resources \cite{zapotoczny2019}, while the latter requires careful selection of aggregation methods \cite{Oneata2021}. 

The evaluation of confidence estimation methods for end-to-end ASR systems is also not trivial. Many widely used in the past confidence estimation metrics, like Area Under the Curve of the Receiver Operating Characteristic ($\mathrm{AUC}_\mathrm{ROC}$) or Area Under the Precision-Recall Curve 
($\mathrm{AUC}_\mathrm{PR}$), produce unreasonably high scores induced by high accuracy of modern ASR with heavy imbalance between correctly and incorrectly recognized words. Also, most of the metrics do not express the possibility of obtaining different results for different values of the confidence threshold of the confidence estimator (the adjustability).

Our main contributions are the following:

\begin{enumerate}
    \vspace{-6pt} 
    \item We propose a fast non-trainable confidence measure based on an exponentially normalized entropy over per-frame probabilities for output tokens.
    \vspace{-6pt} 
    \item We show how per-frame confidence measures can be aggregated to obtain confidence measure per unit and per word for greedy decoding for CTC and RNN-T-based models.
    \vspace{-6pt} 
    \item We evaluate the proposed methods against the traditional method based on probability of the most likely prediction using state-of-the-art Conformer model  and demonstrate that proposed measures are more accurate both for clean and noisy acoustic conditions.
    \vspace{-6pt} 
\end{enumerate}
The implementation of the proposed confidence estimation methods is available in the NeMo toolkit\footnote{\url{https://github.com/NVIDIA/NeMo}}~\cite{kuchaiev2019nemo}.

\section{Entropy-based confidence measures}
\label{sec:entropy_measures}

Consider an end-to-end ASR model which predicts an output unit $u$ based on a probability distribution $p(v)$ over all possible tokens $v$ from a vocabulary $\mathcal{V}$ of size $V$. Let's define a confidence estimator for recognized unit $u$ as mapping $F: p(.) \rightarrow [0,1]$. 
For example, we can estimate confidence using the probability for the most-likely symbol: $F(p) = \max \limits_{v \in \mathcal{V}} p_v$. The minimum of $F(p)$ over all possible distributions $p(.)$ is $(\frac{1}{V})$. The \textit{normalized maximum probability confidence} is defined as follows:
\begin{equation}
\label{equation:max_prob}
    F_{max}(p) = \frac    {F(p) - 1/V}{1 - 1/V} 
    =         \frac{\max p_v - 1/V}{1 - 1/V}
\end{equation}
Recently  Oneaţă et al. \cite{Oneata2021} proposed to estimate confidence with the negative Gibbs entropy:
\begin{equation*} 
  F(p) =  - H_{g}(p) = \sum p_v \ln(p_v)
\end{equation*}
This idea\footnote{Similar confidence measures were used to estimate confidence in other machine learning tasks \cite{tornetta_2021, phan2022sleeptransformer}.} can be reformulated to the \textit{linearly normalized Gibbs entropy-based confidence} through the base-$V$ Shannon entropy $H_{sh}(p)=-\sum p_v \log_V (p_v)$:
\begin{equation} 
\label{equation:lin_ent_gibbs}
\begin{split}
 & F_{g}(p) =  1 - H_{sh}(p) = 
    1 + \sum p_v \log_V (p_v)\\
 &= 1 + \frac{1}{\ln V } \sum p_v \ln(p_v) 
  = 1 - \frac{H_{g}(p)}{\max H_g(p)}
\end{split}
\end{equation}

We will define two new parametric confidence measures using the Tsallis and R{\'e}nyi entropies instead of the Gibbs entropy. The \textit {linearly normalized Tsallis entropy-based confidence} in defined in the following way:
\begin{equation} \label{equation:lin_ent_tsallis}
 F_{ts}(p) = 1 - \frac{H_{ts}(p)}{\max H_{ts}(p)} 
    = \frac{V^{1 - \alpha} - \sum p_v^{\alpha}}{V^{1 - \alpha} - 1}
\end{equation}
where $H_{ts}(p)$ is Tsallis entropy \cite{tsallis1988possible}: 
\begin{equation} \label{equation:ent_ts}
    H_{ts }(p) = \frac{k}{\alpha - 1}
    \Bigl(1 - \sum p_v^{\alpha}\Bigl)
\end{equation}
Here we set $k=1$. The Tsallis entropy becomes the Gibbs entropy when $\alpha \rightarrow 1$. When $0 < \alpha < 1$, it works similar to temperature scaling \cite{hinton2015} where logits are divided by a learned scalar parameter to soften the softmax output. Thus, the Tsallis entropy allows for sensitivity control: smaller $\alpha$ leads to higher sensitivity, which can help mitigating overconfidence.

Similarly, the \textit{linearly normalized R{\'e}nyi entropy-based confidence} in defined as follows:
\begin{equation} \label{equation:lin_ent_renui}
    F_{r}(p) = 1 - \frac{H_{r}(p)}{\max H_{r}(p)} 
    = 1 + \frac{\log_V \bigl(\sum p_v^{\alpha} \bigl)}{\alpha - 1}
\end{equation}
where $H_{r}(p)$ is the R{\'e}nyi entropy \cite{renyi1961measures}:
\begin{equation} \label{equation:ent_renui}
    H_{r}(p) = \frac{1}{\alpha - 1}
    \log_2 \Bigl( \sum p_v^{\alpha}\Bigl) 
\end{equation}
It becomes the base-$2$ Shannon entropy when $\alpha \rightarrow 1$.

\subsection{Entropy measures with exponential normalization}
Instead of linear entropy normalization, we propose to use exponential entropy and normalize it the same way as for the maximum probability. The general formula for the \textit{exponentially normalized entropy-based confidence} is defined as follows:
\vspace{-4pt}
\begin{equation} 
\label{equation:exp_ent}
        F^{e}(p) = \frac{e^{-H(p)} - e^{-\max H(p)}}
        {1 - e^{-\max H(p)}},
\end{equation}
Substituting  the Gibbs, Tsallis and R{\'e}nyi entropies $H_{g}(p)$, $H_{ts}(p)$, and $H_{r}(p)$ in the formula above, we define the following exponential confidence measures:
\begin{equation}
\label{equation:exp_ent_gibbs}
   F^{e}_{g}(p) = \frac{V \cdot e^{(\sum p_v \ln(p_v))} - 1}{V - 1},
\end{equation}
\begin{equation} \label{equation:exp_ent_tsallis}
  F^{e}_{ts}(p) = \frac
  {e^{\frac{1}{1 - \alpha} (V^{1 - \alpha} - \sum p_v^{\alpha})} - 1}
  {e^{\frac{1}{1 - \alpha} (V^{1 - \alpha} - 1)} - 1},
\end{equation}
\begin{equation} \label{equation:exp_ent_renui}
    F^{e}_{r}(p) = \frac{V \bigl(\sum p_v^{\alpha}\bigl)^{\frac{1}{\alpha - 1}} - 1}{V - 1},
\end{equation}
where the R{\'e}nyi exponential normalization is performed with base $2$ to match the logarithm.

\begin{figure*}[t]
\vspace{-15pt}
     \centering
     \begin{subfigure}[t]{0.342\textwidth}
         \centering
         \includegraphics[width=\textwidth]{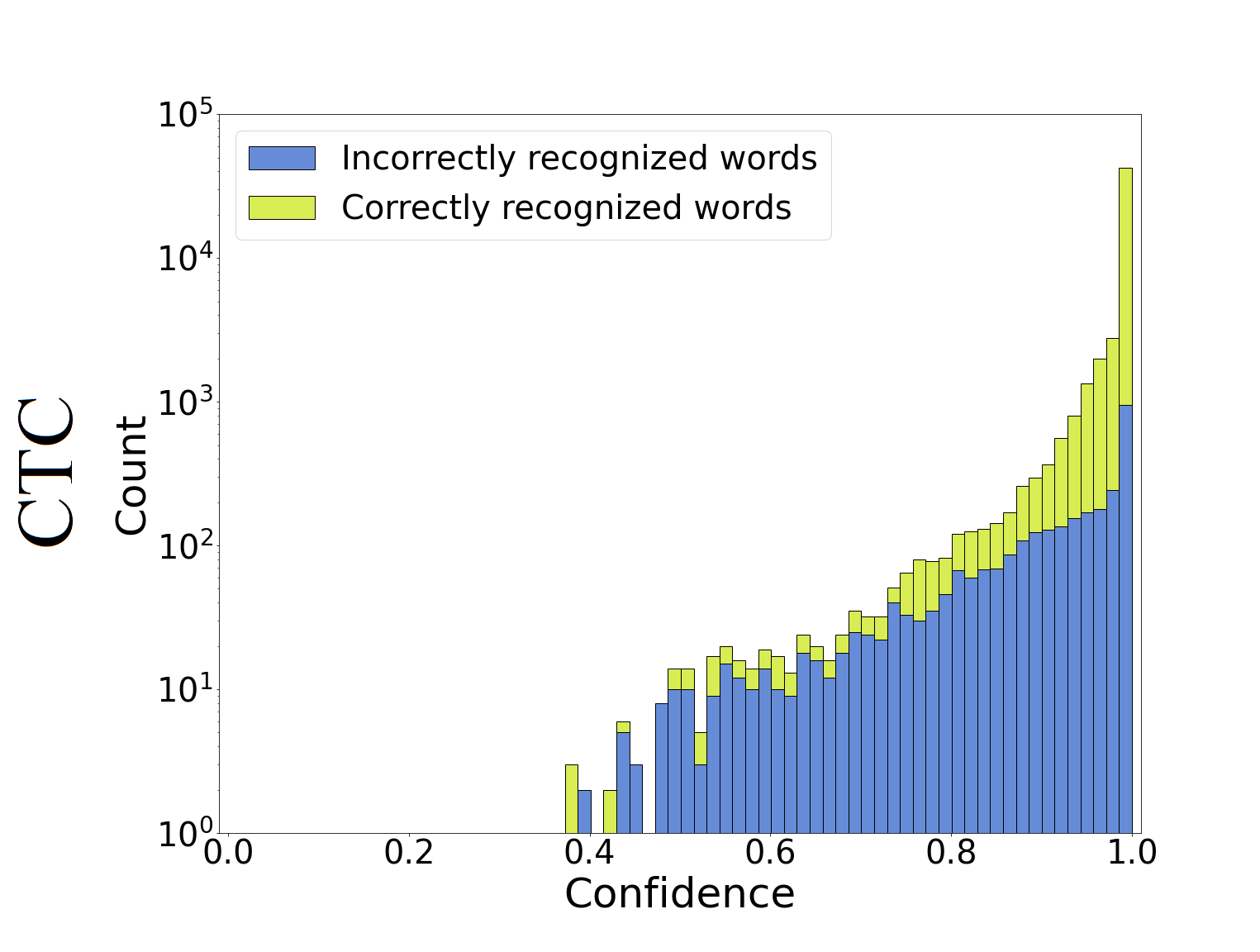}
         \vspace{-20pt}
         \caption{$\operatorname*{mean} F_{max}(p)$}
         \label{fig:hist-ctc-max-prob-mean}
     \end{subfigure}
     \hfill
     \begin{subfigure}[t]{0.324\textwidth}
         \centering
         \includegraphics[width=\textwidth]{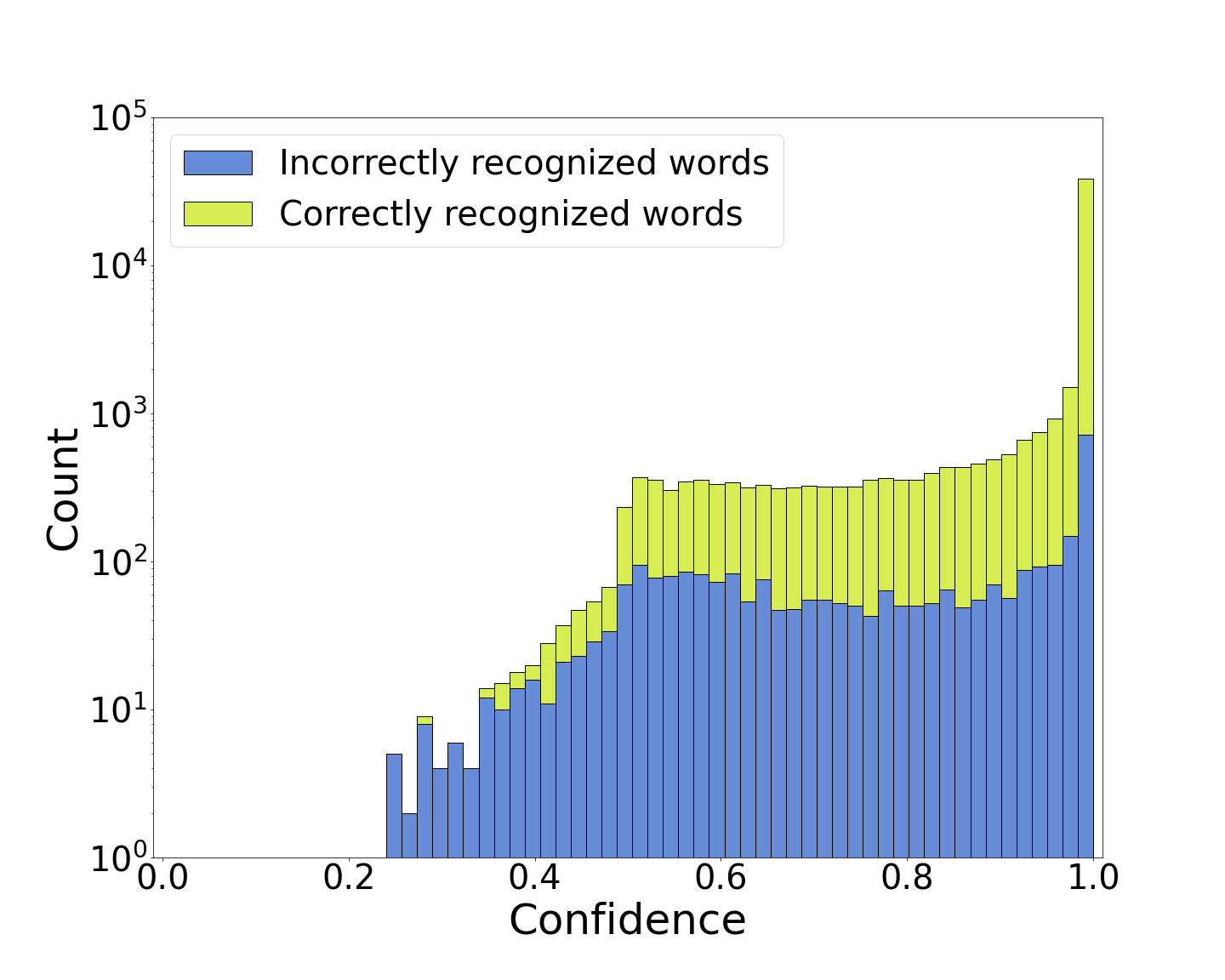}
         \vspace{-20pt}
         \caption{$\min F_{max}(p)$}
         \label{fig:hist-ctc-max-prob-min}
     \end{subfigure}
     \hfill
     \begin{subfigure}[t]{0.324\textwidth}
         \centering
         \includegraphics[width=\textwidth]{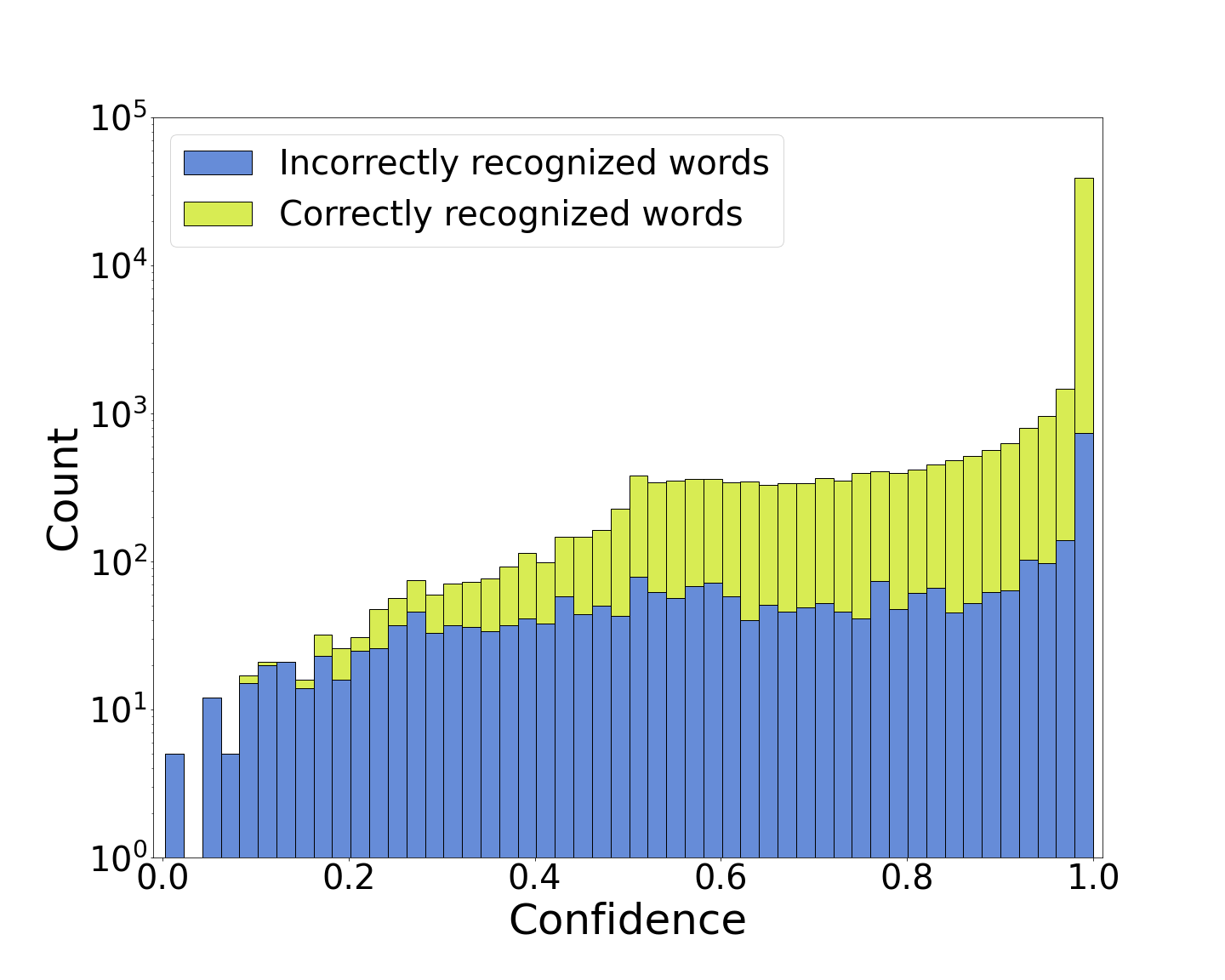}
         \vspace{-20pt}
         \caption{$\prod F_{max}(p)$}
         \label{fig:hist-ctc-max-prob-prod}
     \end{subfigure}
     \hfill
     \begin{subfigure}[t]{0.342\textwidth}
         \centering
         \includegraphics[width=\textwidth]{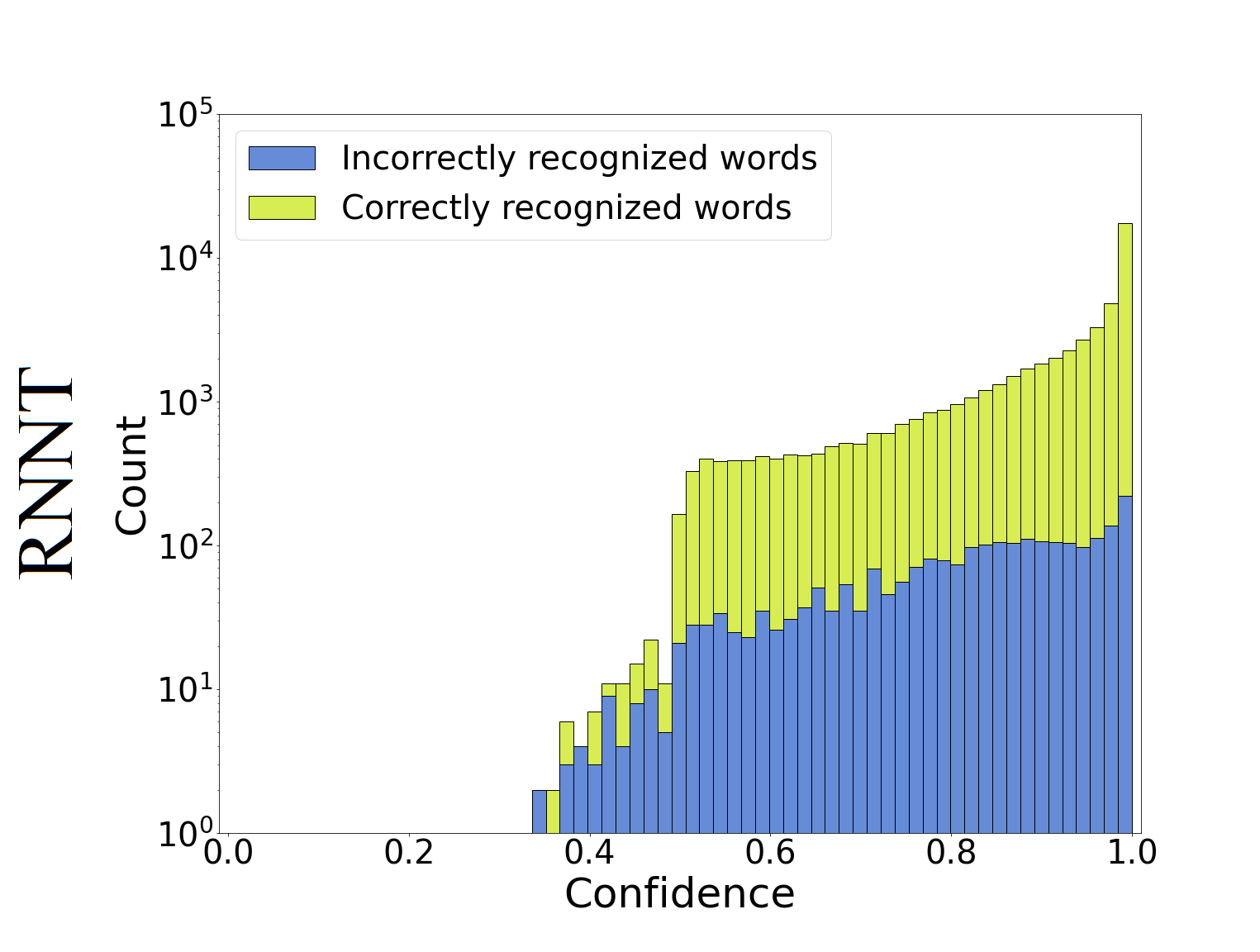}
         \vspace{-20pt}
         \caption{$\operatorname*{mean} F_{max}(p)$}
         \label{fig:hist-rnnt-max-prob-mean}
     \end{subfigure}
     \hfill
     \begin{subfigure}[t]{0.324\textwidth}
         \centering
         \includegraphics[width=\textwidth]{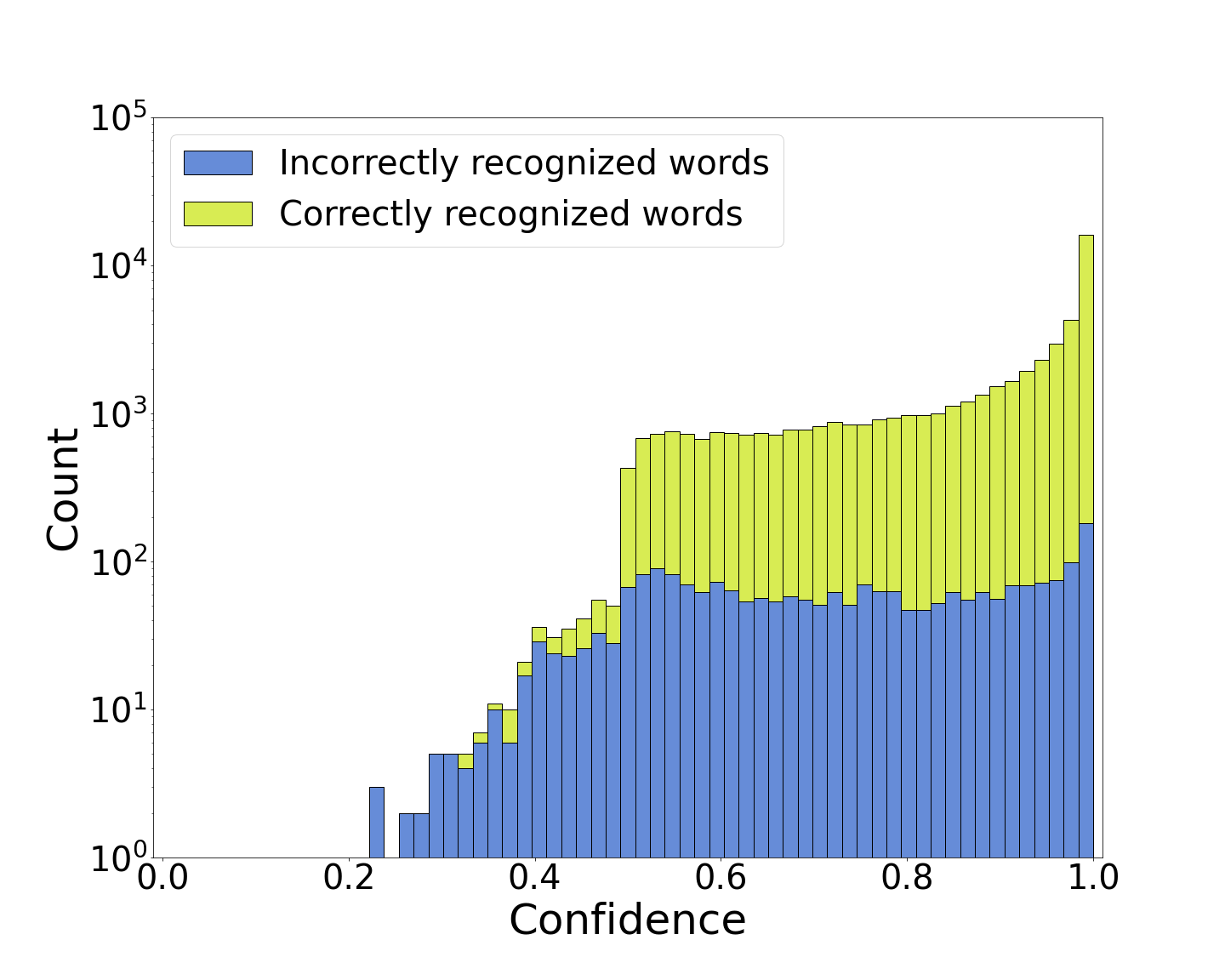}
         \vspace{-20pt}
         \caption{$\min F_{max}(p)$}
         \label{fig:hist-rnnt-max-prob-min}
     \end{subfigure}
     \hfill
     \begin{subfigure}[t]{0.324\textwidth}
         \centering
         \includegraphics[width=\textwidth]{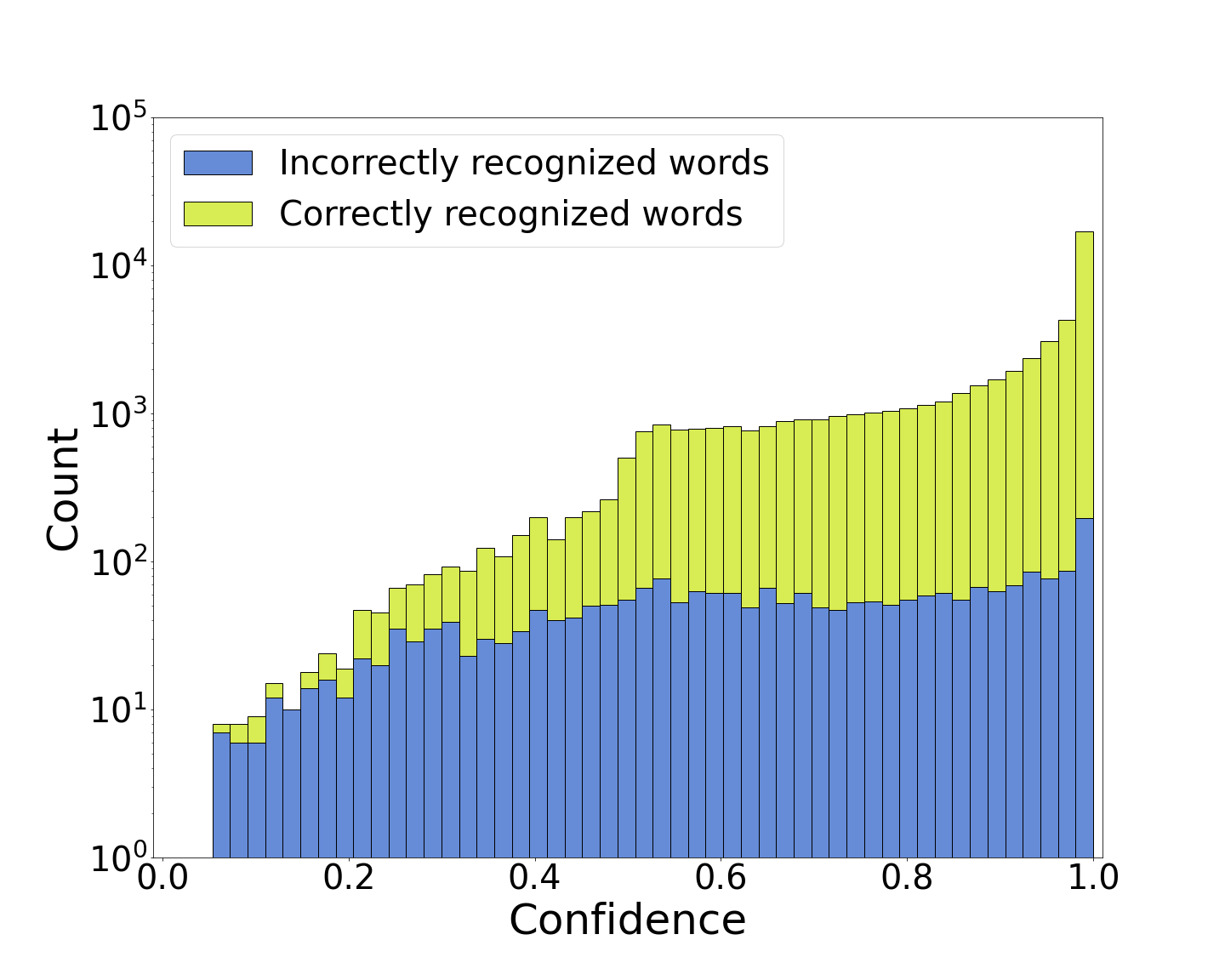}
         \vspace{-20pt}
         \caption{$\prod F_{max}(p)$}
         \label{fig:hist-rnnt-max-prob-prod}
     \end{subfigure}
     \vspace{-5pt} 
    \caption{Stacked log-histograms of correctly and incorrectly recognized words against their maximum-probability confidence scores. The word confidence computed with following aggregation functions: \ref{fig:hist-ctc-max-prob-mean} and \ref{fig:hist-rnnt-max-prob-mean} - $mean$, \ref{fig:hist-ctc-max-prob-min} and \ref{fig:hist-rnnt-max-prob-min} - $minimum$, \ref{fig:hist-ctc-max-prob-prod} and \ref{fig:hist-rnnt-max-prob-prod}- $product$.
    Both models, Conformer-CTC-Large and Conformer-RNN-T-Large, were trained on Librispeech, and tested on Librispeech \textit{test-other}.
    }
\label{fig:hist-max-prob}
\vspace{-10pt} 
\end{figure*}

\section{Word level confidence estimation}

A word-level confidence estimator requires applying additional aggregation over all word units $u_i$ from word $w$:
\begin{equation} \label{equation:word_conf}
q(w) = \operatorname*{agg}\limits_{u_i \in w} q(u_i)
\vspace{-4pt}
\end{equation}
For this work, we use three prediction aggregation functions: arithmetic mean, minimum, and product.

Both CTC and RNN-T ASR models use the special $\langle blank \rangle$ token besides the model vocabulary. One can ignore this token or treat it the same way as other tokens during world-level confidence calculation. We decided to exclude the $\langle blank \rangle$ units from the token-level prediction aggregation, otherwise the scores would be contaminated with information from the deleted units. Without $\langle blank \rangle$ units, unit-level aggregation functions become trivial for RNN-T models and require only a few probability vectors that correspond to consecutive non-$\langle blank \rangle$ units in the CTC case. We use the same token-level and word-level aggregations function for CTC models.


Figure~\ref{fig:hist-max-prob} shows examples of the word-confidence distributions of correctly and incorrectly recognized words for different aggregation functions for the max-probability based confidence $F_{max}(p)$. As one can see, both CTC and RNN-T predictions are overconfident, resulting in an incomplete confidence spectrum coverage for the $mean$- and $minimum$ aggregations. The $product$ aggregation pushes confidence scores of incorrect words towards zero slightly stronger than ones of correct words, reaching the minimum confidence for some of the errors. However, the considered confidence aggregation functions do little to separate the distribution of correct predictions from incorrect ones.



\section{Experimental setup}
\label{sec:setup}

\subsection{Models and data}

We evaluated our confidence estimation methods using the Conformer-CTC and Conformer-Transducer\cite{Gulati2020} models, trained on the LibriSpeech \cite{panayotov2015librispeech} dataset in the NeMo toolkit
\footnote{\url{https://catalog.ngc.nvidia.com/orgs/nvidia/teams/nemo/models/stt_en_conformer_ctc_large_ls},
 \url{https://catalog.ngc.nvidia.com/orgs/nvidia/teams/nemo/models/stt_en_conformer_transducer_large_ls}
}. The Conformer-CTC model has greedy Word Error Rate (WER) 2.7 and 6.1, and Conformer-Transducer - greedy WER 2.3 and 5.1 for \textit{test-clean} and \textit{test-other}, respectively. Models' vocabulary are 128 Byte-Pair Encoding (BPE) wordpieces for Conformer-CTC and 1024 for Conformer-Transducer. These vocabularies give an average of 3 and 1.8 units per word, respectively.

To evaluate confidence under noisy conditions, we added noise from the Freesound database \cite{font2013freesound} and the MUSAN corpus \cite{snyder2015musan}, to the \textit{test-other} set with the following signal-to-noise ratio's (SNR): 0, 10, 20 and 30 dB. After decoding, the recognition results were aligned against the reference texts. Then all deletions were removed because our methods cannot produce confidence for not-emitted units, and the words were classified as correct and incorrect (substitutions and insertions).

To assess the ability of estimators to detect ASR insertions (``hallucinating'') we decoded the ``pure'' strong noise dataset. For this data, Conformer-CTC and Conformer-Transducer got 0.4025 and 0.4275 Word Insertions per Second (WIS).

\subsection{Evaluation metrics}

We used several metrics to evaluate the effectiveness of our confidence methods. Apart from $\mathrm{AUC}_\mathrm{ROC}$ and $\mathrm{AUC}_\mathrm{PR}$, we used $\mathrm{AUC}_\mathrm{NT}$, which represents the area under the $\mathrm{NPV}\sim\mathrm{TNR}$ curve (Negative Predictive Value vs. True Negative Rate) and demonstrates the effectiveness of detecting erroneously transcribed words. This metric can be thought of as an $\mathrm{AUC}_\mathrm{PR}$ in which errors are treated as positives.

The $\mathrm{AUC}$-based metrics do not take into account the class imbalance of correctly and incorrectly recognized words, so we considered additional metrics to make more informed decisions on the choice of the best performing confidence methods.

Normalized Cross Entropy ($\mathrm{NCE}$) and Expected Calibration Error ($\mathrm{ECE}$) measure how close the correct word confidence scores are to $1.0$ and the incorrect word scores are to $0.0$. $\mathrm{NCE}$ (also known as the Normalized Mutual Information \cite{siu97_eurospeech}) ranges from $-\infty$ to $1.0$, with negative scores indicating that the conﬁdence method performs worse than the setting confidence score to $1-\mathrm{WER}$. One downside of relying on this metric is that it favors overconfident estimation methods if the model WER is low. $\mathrm{ECE}$ \cite{naeini2015obtaining} ranges from $0.0$ to $1.0$ with the best value $0.0$. 

\vspace{-2pt} 
\subsection{Proposed evaluation metrics}
\vspace{-2pt} 
We believe that the values of the confidence estimator should not be clamped near zero and one, but instead should represent the entire spectrum of probabilities. We call this property of a confidence estimator adjustability (responsiveness), i.e. the ability to produce different classification results for different threshold values. We propose to evaluate the adjustability and general usability of confidence estimation methods using the following metrics.

Youden's curve \cite{youden1950} (or Youden's J statistic) is a way of summarising the performance of a diagnostic test, which is often used in conjunction with the $\mathrm{ROC}$ analysis. One of its advantages is its insensitivity to class imbalance. For confidence evaluation, we define the Youden's curve statistics as follows:
\begin{equation} \label{equation:yc}
\vspace{-5pt} 
    \mathrm{STAT}_\mathrm{YC} = \operatorname*{STAT}\limits_{\tau \in [0,1]}(\mathrm{TNR(\tau)} - \mathrm{FNR(\tau)}),
\end{equation}
where $\mathrm{STAT}$ is the desired statistic, $\mathrm{YC}$ is the Youden's curve, and $\mathrm{TNR(\tau)}$ and $\mathrm{FNR(\tau)}$ are True Negative and False Negative Rates with respect to the threshold $\tau$.

We propose measuring area under the curve $\mathrm{AUC}_\mathrm{YC} \in [0,1]$ with a desired value around $0.5$, the maximum value $\mathrm{MAX}_\mathrm{YC} \in [0,1]$ with a desired value close to $1.0$, and the standard deviation of values $\mathrm{STD}_\mathrm{YC} \in [0,0.5]$ with a desired value around $0.25$. $\mathrm{AUC}_\mathrm{YC}$ represents the rate of the effective threshold range (i.e. the adjustability). $\mathrm{MAX}_\mathrm{YC}$ shows the optimal $\mathrm{TNR}$ vs. $\mathrm{FNR}$ tradeoff and used as a criterion for selecting the optimal $\tau$. $\mathrm{STD}_\mathrm{YC}$ is a supplementary metric that indicates that $\mathrm{TNR}$ and $\mathrm{FNR}$ increase at different rates as the $\tau$ increases. If $\mathrm{AUC}_\mathrm{YC} < \mathrm{STD}_\mathrm{YC}$, then the confidence spectrum coverage is most likely incomplete.

We also measured $\mathrm{TNR}_{.05}$ as $\mathrm{TNR}(Y,\tau): \mathrm{FNR}(X,\tau) \approx 0.05$ ($X$ and $Y$ are non-overlapping datasets) to evaluate the hallucinations reduction ability of an estimator. The \textit{noise} data does not have positive examples, so we used the \textit{test-other} test set to choose the optimal $\tau$ and then measured $\mathrm{TNR}_{.05}$ on the \textit{noise} data.

\section{Experimental results}
\label{sec:results}

\begin{table}[!t]
\renewcommand*{\arraystretch}{1.1}
\center
\small
\setlength{\tabcolsep}{2pt}
\caption{
 Conformer-CTC: evaluation of confidence methods: $\prod F_{max}(p)$, $\prod F^{e}_{g}(p)$, $\operatorname*{mean} F^{e}_{ts}(p)$, and $\min F^{e}_{ts}(p)$ with metrics: 
 $\mathrm{AUC}_\mathrm{ROC}$, $\mathrm{AUC}_\mathrm{PR}$, $\mathrm{AUC}_\mathrm{NT}$, $\mathrm{AUC}_\mathrm{YC}$ and $\mathrm{TNR}_{.05}$  in $\%$.  The last row is a measure from \cite{Oneata2021} which is similar to a not normalized $\prod F_{max}(p)$. LS \textit{test-clean} and \textit{test-other}.}
 \vspace{-5pt} 
\label{tab:main-ctc}
\adjustbox{max width=\linewidth}{
\begin{tabular}{l|cccc|cccc|c}
    \toprule
        \makecell[r]{\textbf{Test set}}
        & \multicolumn{4}{c|}{\textit{test-clean}}
        & \multicolumn{4}{c|}{\textit{test-other}}
        & \textit{noise} \\
        \midrule
        \textbf{Method}
        & $\mathrm{ROC}$ & $\mathrm{PR}$ & $\mathrm{NT}$ & $\mathrm{YC}$
        & $\mathrm{ROC}$ & $\mathrm{PR}$ & $\mathrm{NT}$ & $\mathrm{YC}$
        & \hspace{-1pt}$\mathrm{TNR}_{.05}$ \\
        \midrule
        $\prod F_{max}(p)$   & 82.15   & 99.36   & 14.60   & 19.18   & 84.66   & 98.79   & 32.41   & 23.01   & 36.79   \\
        $\prod F^{e}_{g}(p)$   & 82.70   & 99.37   & 18.95   & 28.97   & 85.44   & 98.85   & 37.46   & 34.60   & \textbf{40.49}   \\
        $\operatorname*{mean} F^{e}_{ts}(p)$   & \textbf{88.56}   & \textbf{99.55}   & \textbf{36.50}   & 25.94   & \textbf{90.34}   & \textbf{99.21}   & 45.18   & 37.36   & 33.39   \\
        $\min F^{e}_{ts}(p)$   & 84.63   & 99.43   & 30.82   & \textbf{38.11}   & 88.04   & 99.04   & \textbf{47.01}   & \textbf{45.86}   & 37.72   \\
        \midrule
        $F_{max}(p)$ \cite{Oneata2021}   & 82.41   & 99.21   & 21.55   & -   &  81.75   & 98.10   & 29.99   & -   & -   \\
 \bottomrule
\end{tabular}
}
\vspace{-5pt} 
\end{table}

\begin{table}[t]
\renewcommand*{\arraystretch}{1.1}
\center
\small
\setlength{\tabcolsep}{2pt}
\caption{ Conformer-RNN-T: evaluation of confidence methods: $\prod F_{max}(p)$, $\prod F^{e}_{g}(p)$, $\operatorname*{mean} F^{e}_{ts}(p)$, and $\min F^{e}_{ts}(p)$ with metrics: 
 $\mathrm{AUC}_\mathrm{ROC}$, $\mathrm{AUC}_\mathrm{PR}$, $\mathrm{AUC}_\mathrm{NT}$, $\mathrm{AUC}_\mathrm{YC}$ and $\mathrm{TNR}_{.05}$. 
 LS \textit{test-clean} and \textit{test-other}.}
 \vspace{-5pt} 
\label{tab:main-rnnt}
\adjustbox{max width=\linewidth}{
\begin{tabular}{l|cccc|cccc|c}
    \toprule
        \makecell[r]{\textbf{Test set}}
        & \multicolumn{4}{c|}{\textit{test-clean}}
        & \multicolumn{4}{c|}{\textit{test-other}}
        & \textit{noise} \\
        \midrule
        \textbf{Method}
        & $\mathrm{ROC}$ & $\mathrm{PR}$ & $\mathrm{NT}$ & $\mathrm{YC}$
        & $\mathrm{ROC}$ & $\mathrm{PR}$ & $\mathrm{NT}$ & $\mathrm{YC}$
        & \hspace{-1pt}$\mathrm{TNR}_{.05}$ \\
        \midrule
        $\prod F_{max}(p)$   & 72.00   & 99.04   & 08.70   & 15.80   & 75.48   & 98.21   & 21.28   & 19.60   & 17.82   \\
        $\prod F^{e}_{g}(p)$   & 74.85   & 99.12   & 13.07   & 23.04   & 78.16   & 98.37   & 27.70   & 26.80   & 20.99   \\
        $\operatorname*{mean} F^{e}_{ts}(p)$   & 72.50   & 99.07   & 19.86   & 16.27   & 79.52   & 98.47   & 32.61   & 23.55   & 32.44   \\
        $\min F^{e}_{ts}(p)$   & \textbf{81.67}   & \textbf{99.34}   & \textbf{37.49}   & \textbf{25.13}   & \textbf{85.85}   & \textbf{98.90}   & \textbf{47.17}   & \textbf{30.71}   & \textbf{38.58}   \\
 \bottomrule
\end{tabular}
}
\vspace{-15pt} 
\end{table}

\begin{figure*}[t]
\vspace{-15pt}
     \centering
     \begin{subfigure}[t]{0.342\textwidth}
         \centering
         \includegraphics[width=\textwidth]{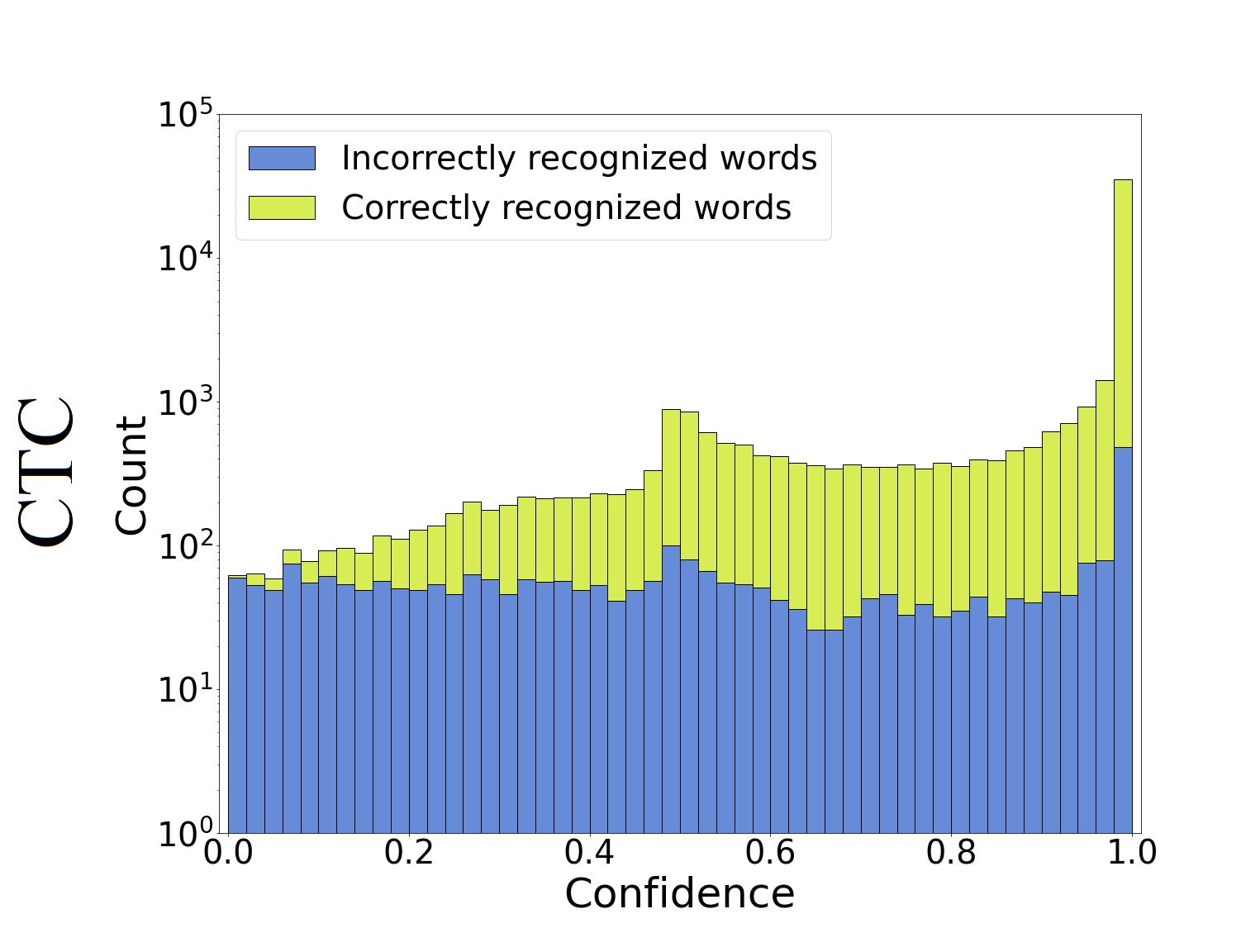}
         \vspace{-20pt}
         \caption{$\prod F^{e}_{g}(p)$}
         \label{fig:hist-ctc-norm-ent-prod-tsallis-exp}
     \end{subfigure}
     \hfill
     \begin{subfigure}[t]{0.324\textwidth}
         \centering
         \includegraphics[width=\textwidth]{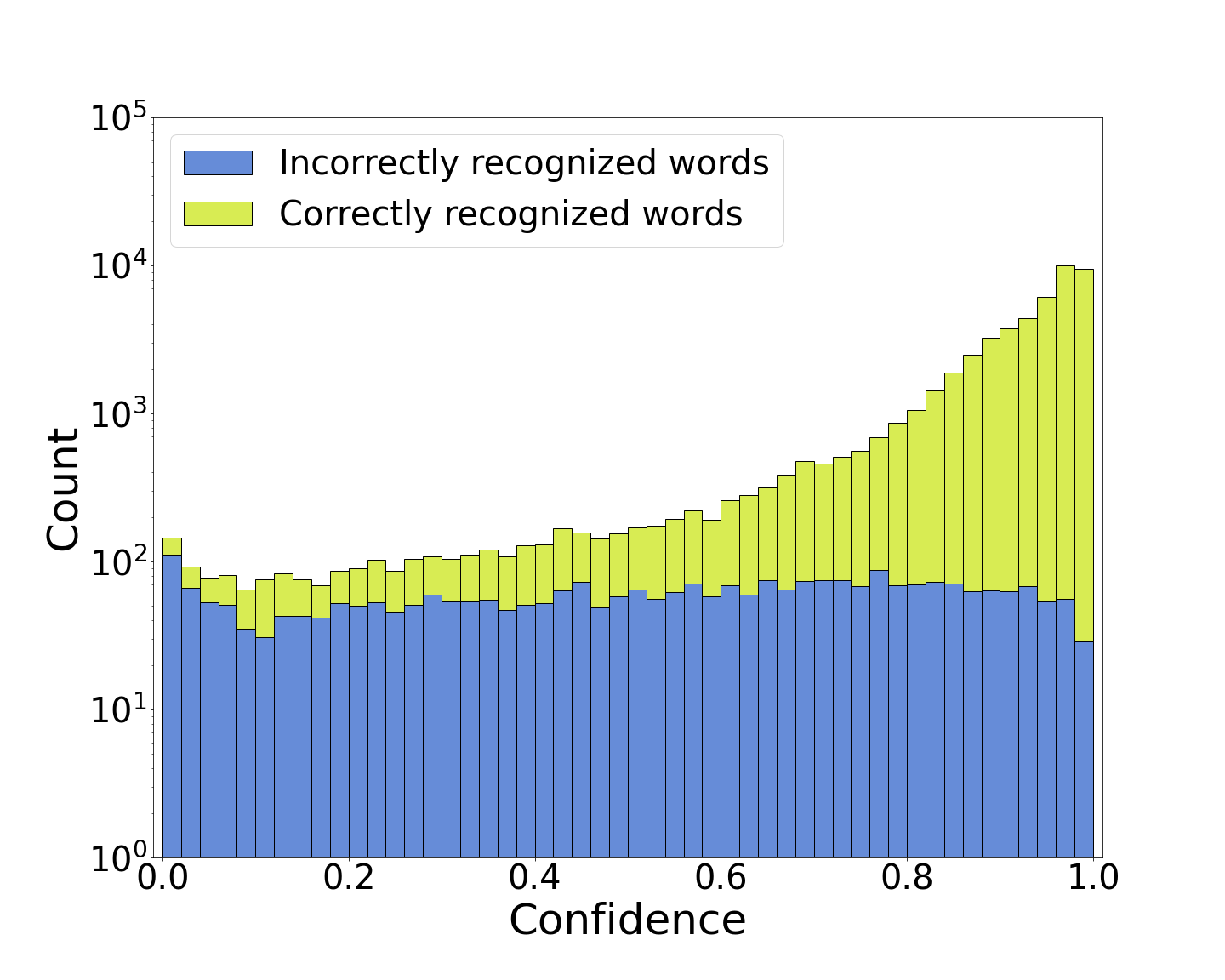}
         \vspace{-20pt}
         \caption{$\operatorname*{mean} F^{e}_{ts}(p)$}
         \label{fig:hist-ctc-norm-ent-mean-tsallis-exp-0.33}
     \end{subfigure}
     \hfill
     \begin{subfigure}[t]{0.324\textwidth}
         \centering
         \includegraphics[width=\textwidth]{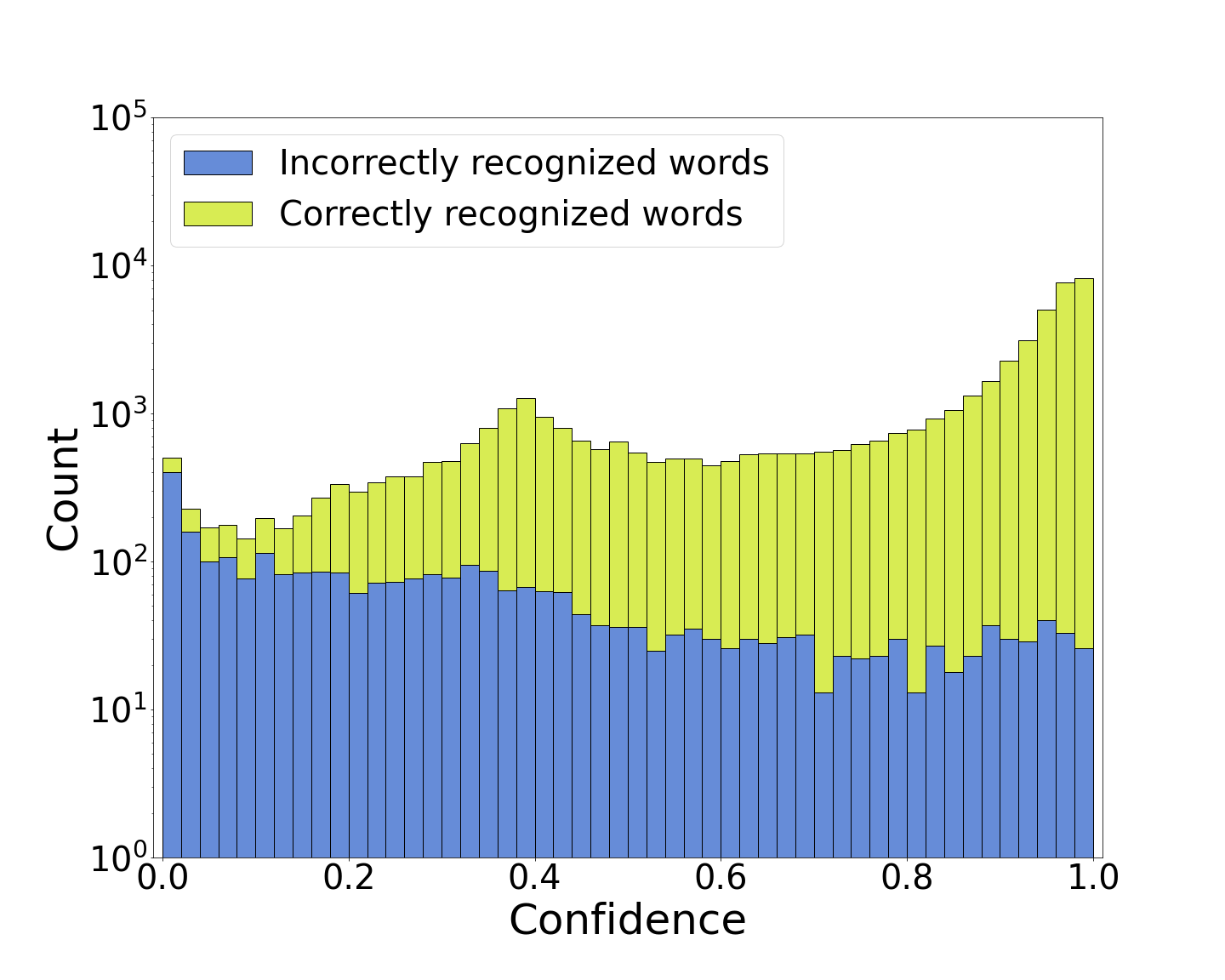}
         \vspace{-20pt}
         \caption{$\min F^{e}_{ts}(p)$}
         \label{fig:hist-ctc-norm-ent-min-tsallis-exp-0.33}
     \end{subfigure}
     \begin{subfigure}[t]{0.342\textwidth}
         \centering
         \includegraphics[width=\textwidth]{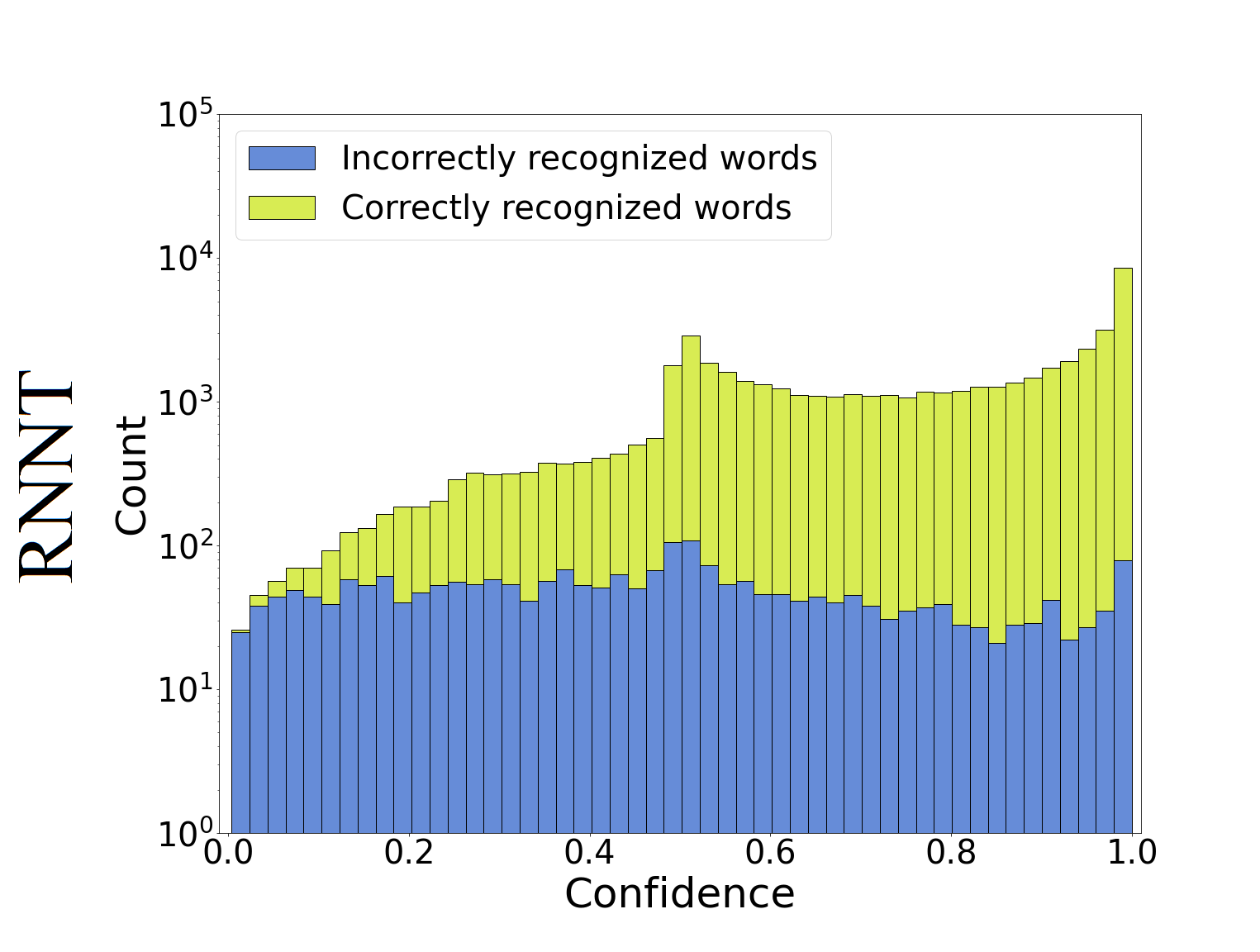}
         \vspace{-20pt}
         \caption{$\prod F^{e}_{g}(p)$}
         \label{fig:hist-rnnt-norm-ent-prod-tsallis-exp}
     \end{subfigure}
     \hfill
     \begin{subfigure}[t]{0.324\textwidth}
         \centering
         \includegraphics[width=\textwidth]{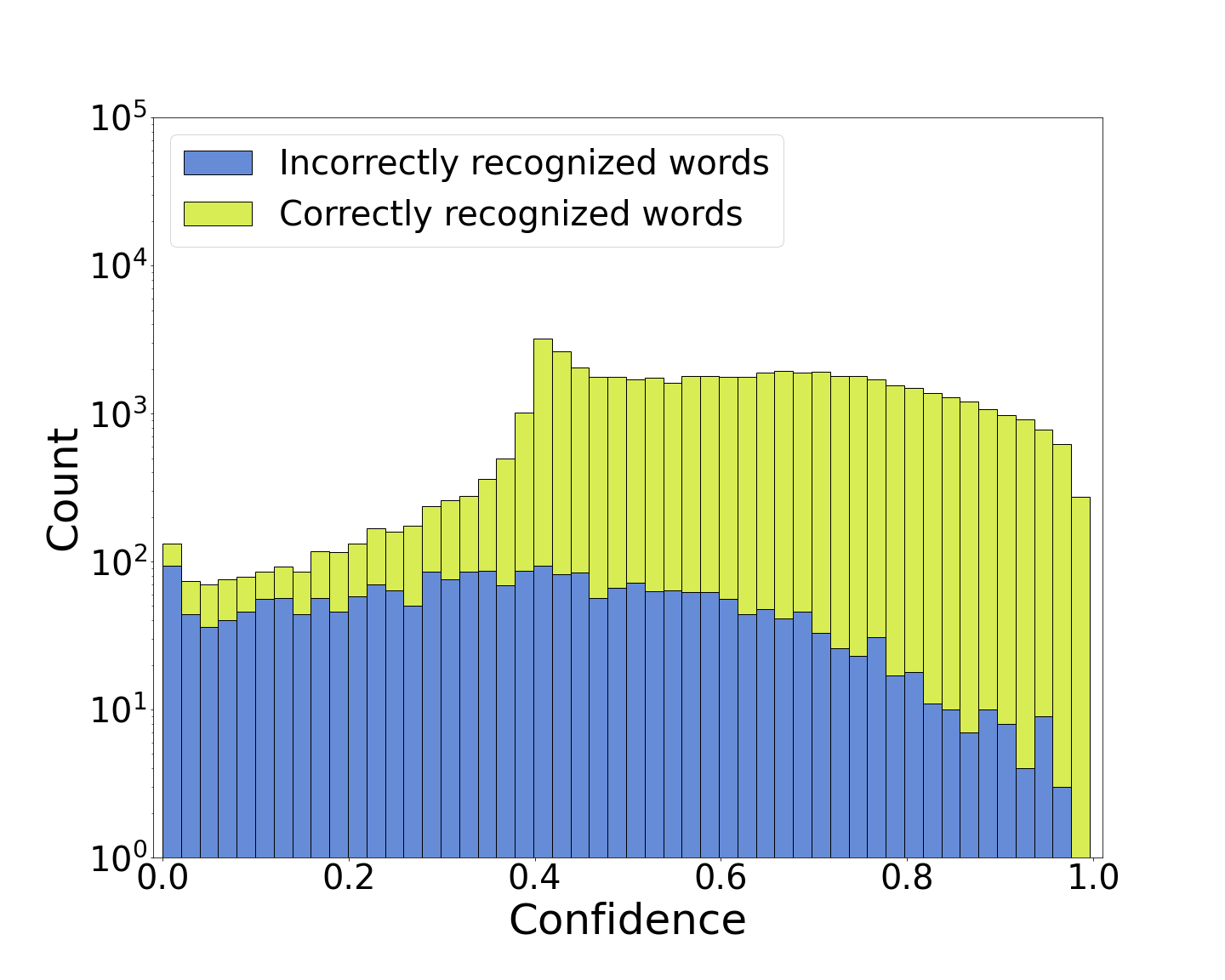}
         \vspace{-20pt}
         \caption{$\operatorname*{mean} F^{e}_{ts}(p)$}
         \label{fig:hist-rnnt-norm-ent-mean-tsallis-exp-0.33}
     \end{subfigure}
     \hfill
     \begin{subfigure}[t]{0.324\textwidth}
         \centering
         \includegraphics[width=\textwidth]{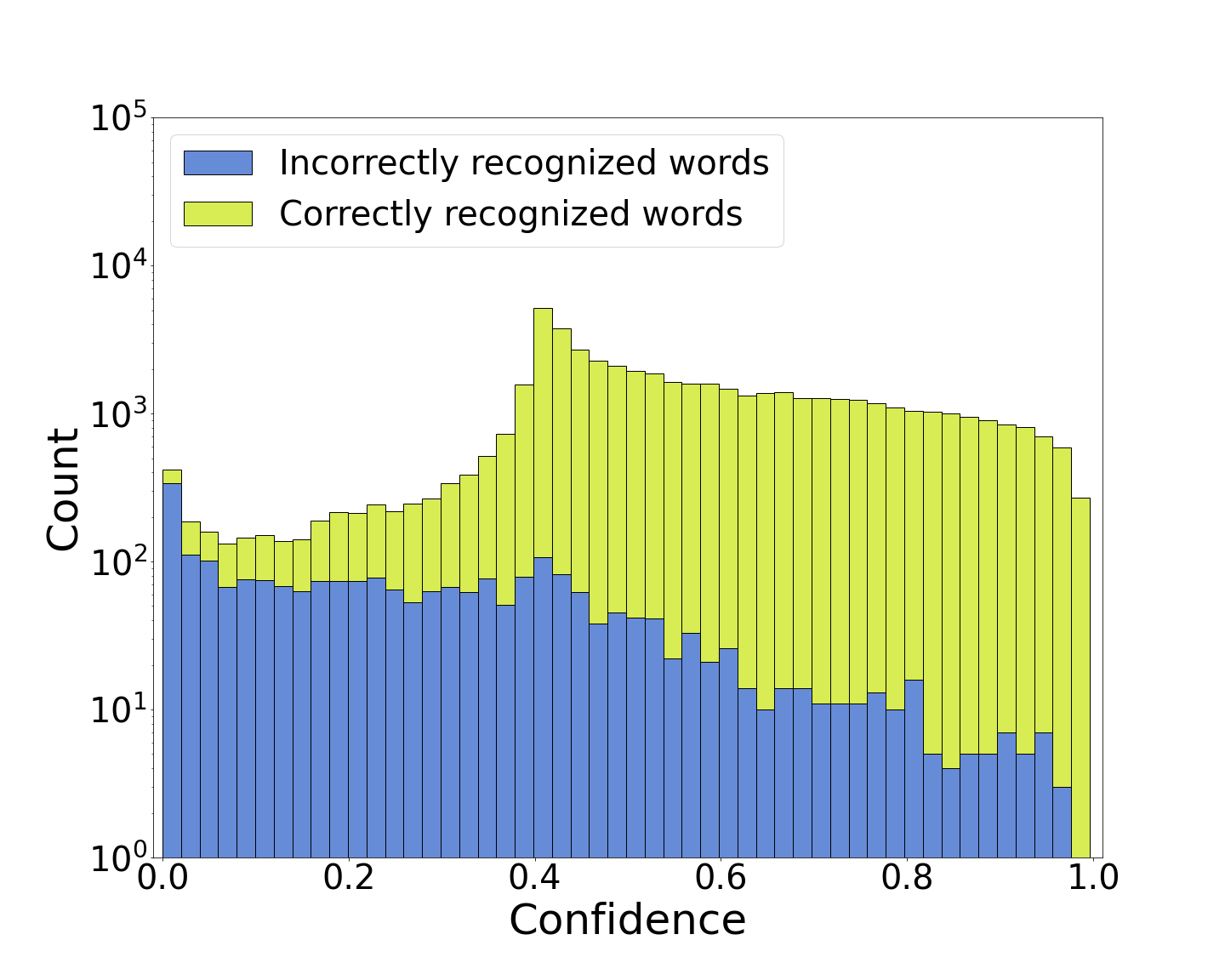}
         \vspace{-20pt}
         \caption{$\min F^{e}_{ts}(p)$}
         \label{fig:hist-rnnt-norm-ent-min-tsallis-exp-0.33}
     \end{subfigure}
     \vspace{-5pt} 
        \caption{Stacked log-histograms of correctly and incorrectly recognized words with respect to their confidence estimation methods. Confidence methods: $\prod F^{e}_{g}(p)$, $\operatorname*{mean} F^{e}_{ts}(p)$, and $\min F^{e}_{ts}(p)$. The Conformer-CTC and Conformer-RNN-T models were trained on LS and tested on \textit{test-other}.}
        \label{fig:hist-best-methods}
\vspace{-20pt} 
\end{figure*}

Tables~\ref{tab:main-ctc}~and~\ref{tab:main-rnnt} show the $\mathrm{AUC}$-based metrics and $\mathrm{TNR}_{.05}$ of the optimal normalized maximum probability, normalized Gibbs entropy, and parametric normalized entropy confidence estimators. Instead of maximizing any particular metric, we chose methods that performed well enough for all metrics in general. Metrics $\mathrm{NCE}$, $\mathrm{ECE}$, $\mathrm{MAX}_\mathrm{YC}$, and $\mathrm{STD}_\mathrm{YC}$ are omitted from the tables for the sake of compactness. We found that the product aggregation is the best choice for the maximum probability and the Gibbs entropy based methods.

The entropy-based methods performed best with the exponential normalization. Among the chosen $1/4$, $1/3$, and $1/2$ $\alpha$ values, $1/3$ delivered the best metric scores for both CTC and RNN-T models. Our parametric estimators $\operatorname*{mean} F^{e}_{ts}(p)$ and $\min F^{e}_{ts}(p)$ are comparable for the CTC system, but for RNN-T $\min F^{e}_{ts}(p)$ is clearly better. Compared to $\prod F_{max}(p)$, $\min F^{e}_{ts}(p)$ is from $1.5$ to $4$ times better at detecting incorrect words (metric $\mathrm{AUC}_\mathrm{NT}$). Thus, we recommend this method as a universal choice for CTC and RNN-T systems.

Table~\ref{tab:main-ctc} also contains results of the most effective measure from \cite{Oneata2021}, which is similar to a not normalized $\prod F_{max}(p)$. Since our CTC and their CTC-Attention systems have similar WER and their confidence measures are similar, it can be concluded that the quality of confidence does not depend on the ASR model type, but on the estimator and WER on the test set.

\input{nt_plot}

Figure~\ref{fig:nt-plot} confirms the conclusion above. The smaller the SNR, the more recognition errors. But with these errors, the model is less certain than usual, resulting in better error detection with a proper confidence measure. Interestingly, the Gibbs entropy performs better than the parametric ones for CTC on the SNR $0$ data. This may indicate that for heavily noise-augmented data the ASR system is unsure enough and further soft-max smoothing is not needed.

Tables~\ref{tab:main-ctc}~and~\ref{tab:main-rnnt}, and Figure~\ref{fig:nt-plot} demonstrate that $\min F^{e}_{ts}(p)$ not only greatly improves the wrong words detection quality ($\mathrm{AUC}_\mathrm{NT}$ and $\mathrm{TNR}_{.05}$) but also makes confidence estimation for RNN-T models almost as effective as for CTC models.

\begin{figure*}[t]
\vspace{-15pt}
     \centering
     \begin{subfigure}[t]{0.245\textwidth}
         \centering
         \includegraphics[width=\textwidth]{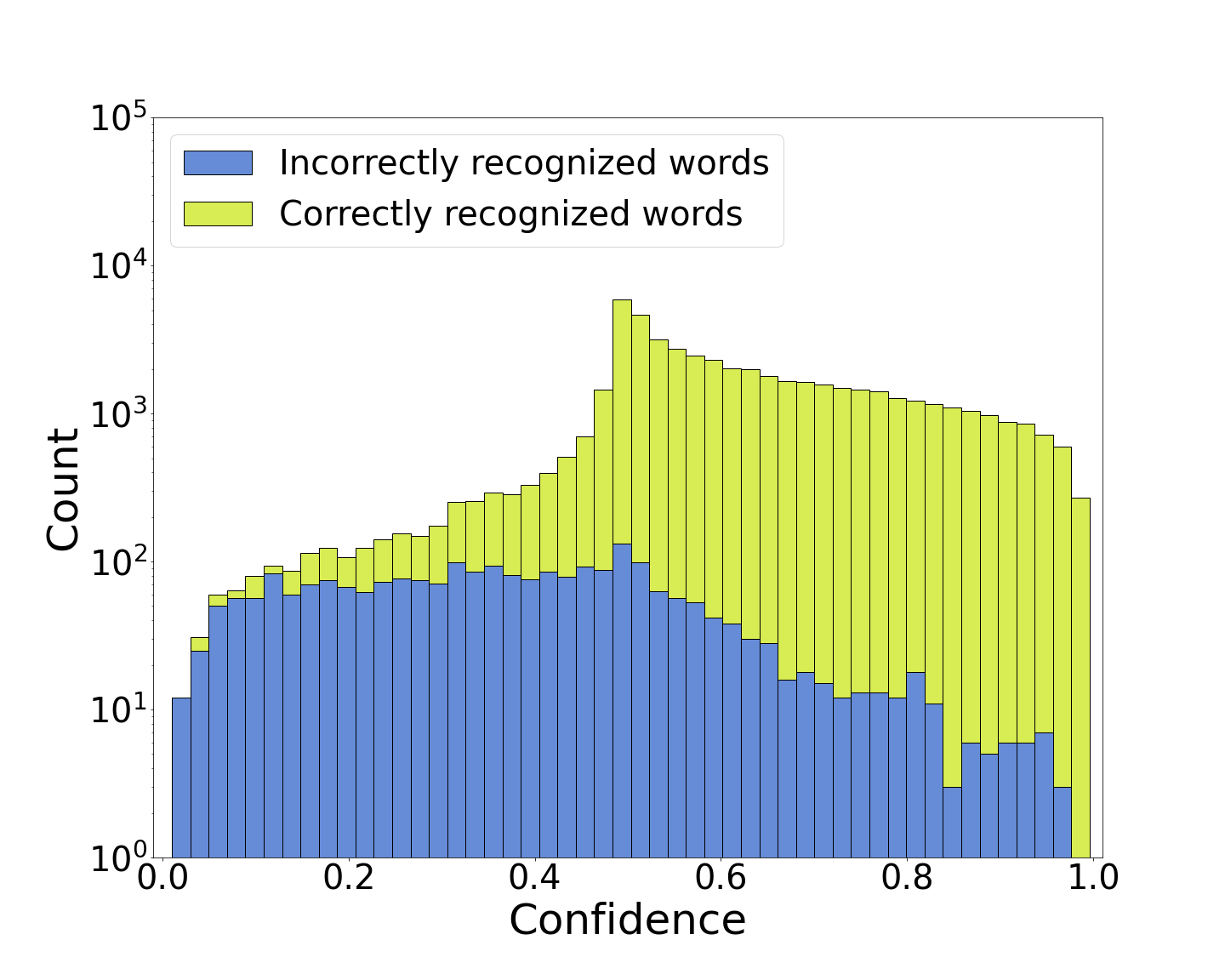}
         \vspace{-15pt}
         \caption{$F^{e}_{r}(p)$, $\alpha=1/3$}
         \label{fig:hist-rnnt-norm-ent-min-renui-exp-0.33}
     \end{subfigure}
     \hfill
     \begin{subfigure}[t]{0.245\textwidth}
         \centering
         \includegraphics[width=\textwidth]{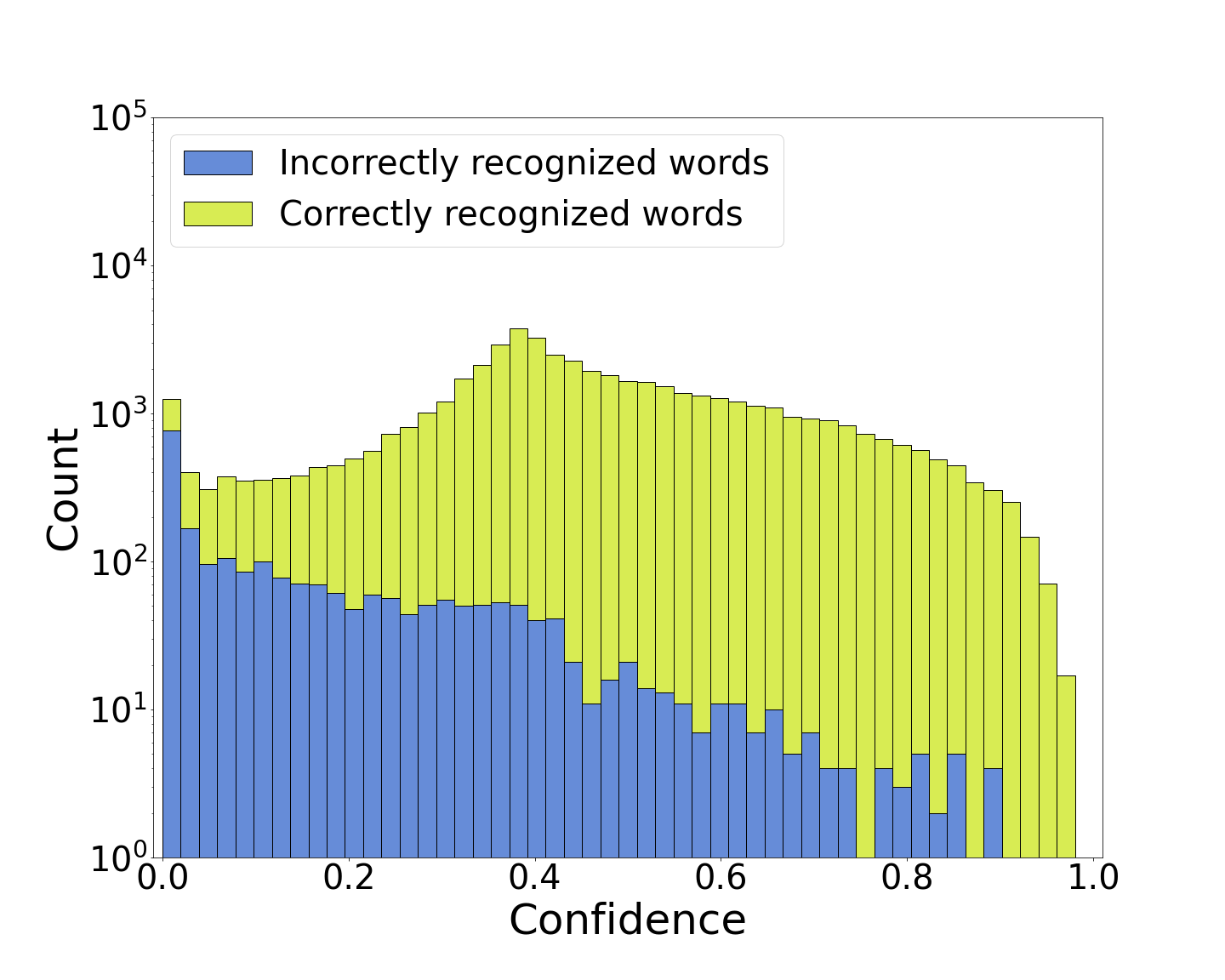}
         \vspace{-15pt}
         \caption{$F^{e}_{ts}(p)$, $\alpha=1/4$}
         \label{fig:hist-rnnt-norm-ent-min-tsallis-exp-0.25}
     \end{subfigure}
     \hfill
     \begin{subfigure}[t]{0.245\textwidth}
         \centering
         \includegraphics[width=\textwidth]{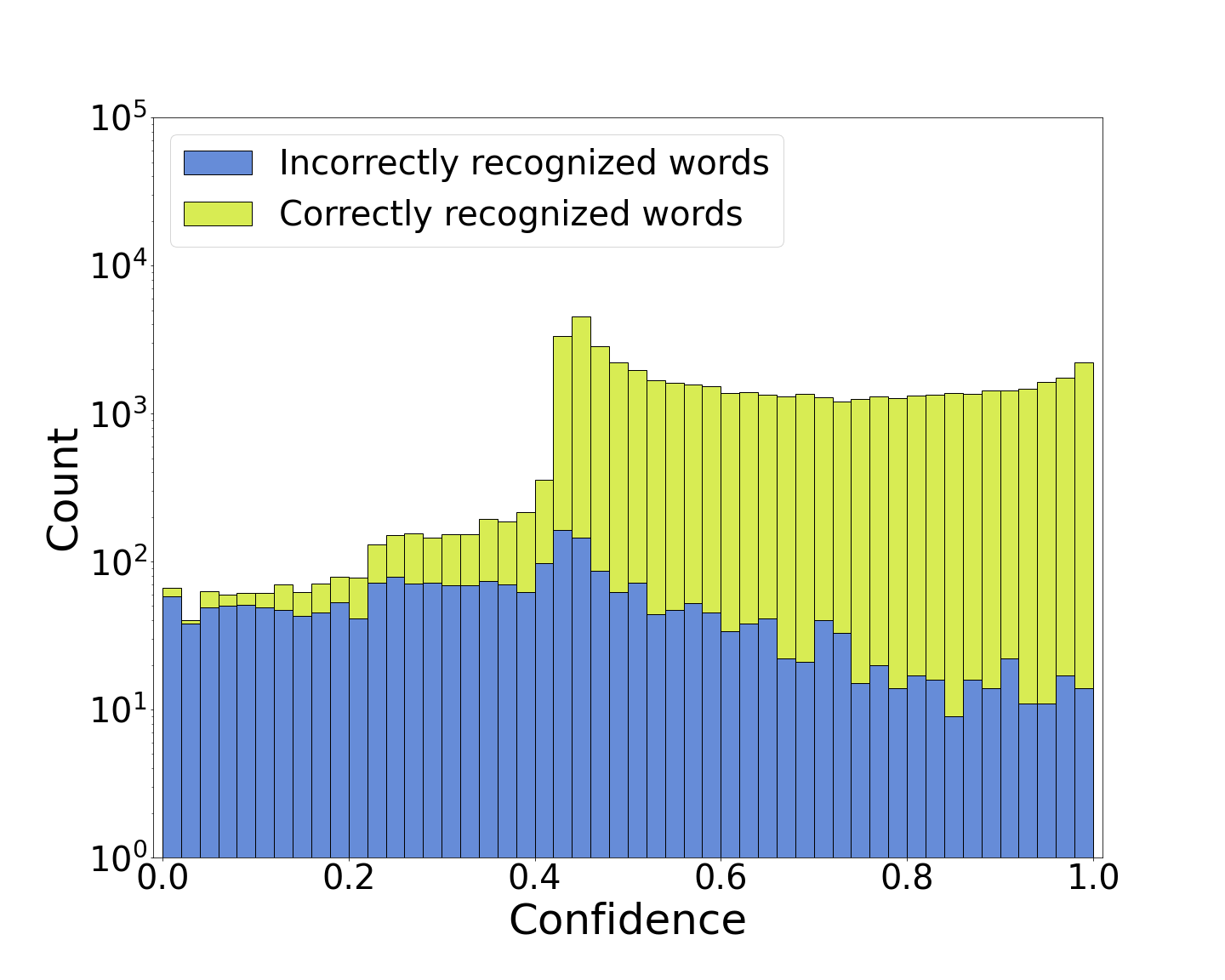}
         \vspace{-15pt}
         \caption{$F^{e}_{ts}(p)$, $\alpha=1/2$}
         \label{fig:hist-rnnt-norm-ent-min-tsallis-exp-0.5}
     \end{subfigure}
     \begin{subfigure}[t]{0.245\textwidth}
         \centering
         \includegraphics[width=\textwidth]{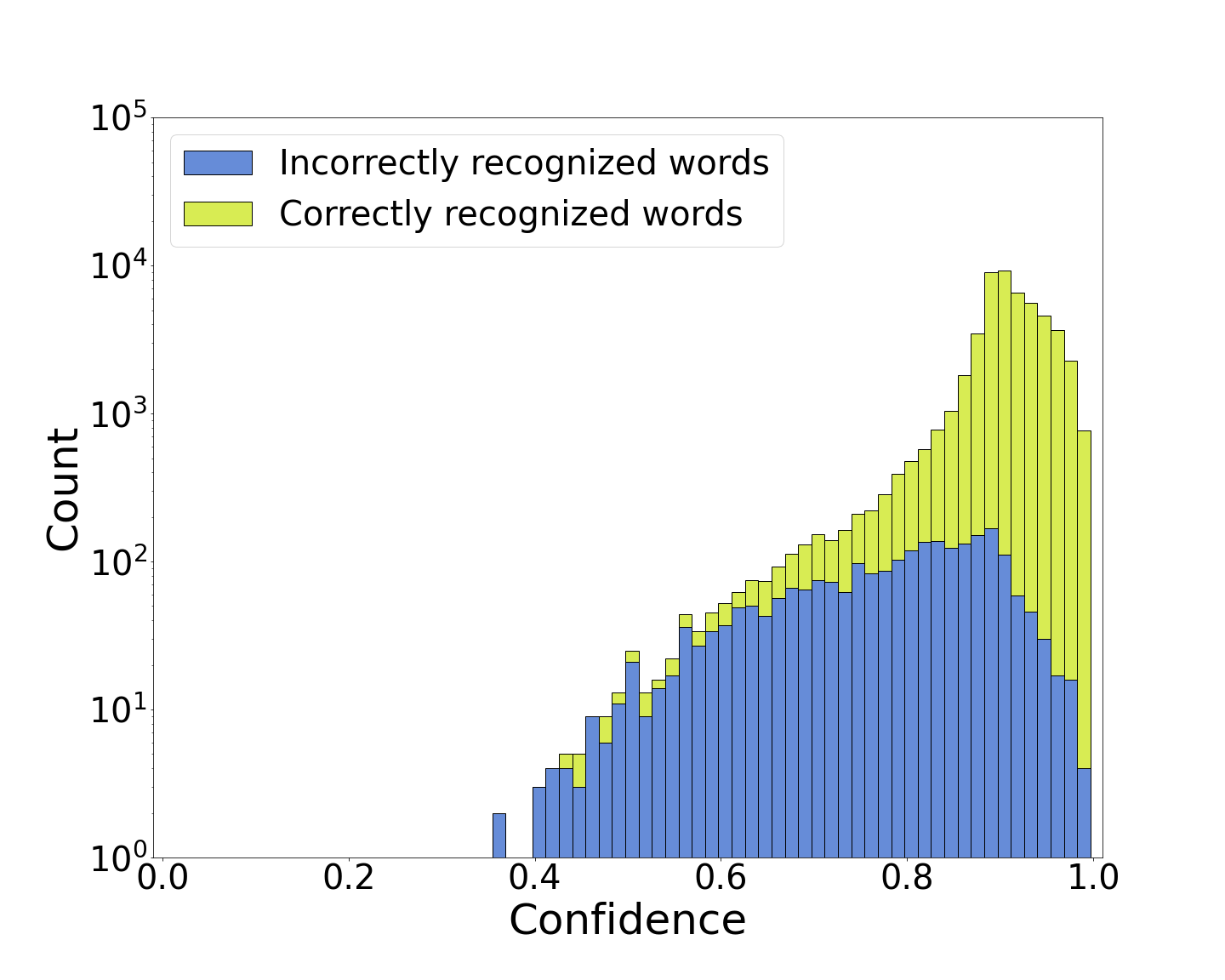}
         \vspace{-15pt}
         \caption{$F_{r}(p)$, $\alpha=1/4$}
         \label{fig:hist-rnnt-norm-ent-min-renui-lin-0.25}
     \end{subfigure}
     \vspace{-5pt} 
        \caption{Stacked log-histograms of correctly and incorrectly recognized words with respect to their confidence measures and $min$ aggregation function. Confidence measures: $F^{e}_{r}(p)$ with $\alpha=1/3$, $F^{e}_{ts}(p)$ with $\alpha=1/4$, $F^{e}_{ts}(p)$ with $\alpha=1/2$, and $F_{r}(p)$ with $\alpha=1/4$. Conformer-RNN-T, evaluated on LS \textit{test-other}.}
        \label{fig:hist-ent-measures}
\vspace{-10pt} 
\end{figure*}

\vspace{15pt} 
\section{Discussion}
\label{sec:discussion}
\vspace{-5pt} 

\begin{table}[!b]
\vspace{-5pt} 
\center
\small
\setlength{\tabcolsep}{2.5pt}
\caption{Percentage of errors and overall number of words against the number of units in word on LibriSpeech \textit{test-clean}.}
\vspace{-5pt} 
\label{tab:units-per-word}
\adjustbox{max width=\linewidth}{
\begin{tabular}{l|cccc|cccc}
    \toprule
    \textbf{Model}                                      &
    \multicolumn{4}{c|}{Conformer-CTC}         &
    \multicolumn{4}{c}{Conformer-RNN-T}        \\
    \midrule
    \textbf{\# of units}    & 1-2   & 3-4   & 5-6   & $>$6   & 1-2   & 3-4   & 5-6   & $>$6   \\
    \midrule
    \textbf{\% of errors}   & 1.1   & 2     & 3.3   & 5.5   & 1     & 3.3   & 4.2   & 7.6   \\
    \textbf{\# of words}    & 25576 & 15316 & 8771  & 2932  & 40310 & 9220  & 2827  & 238   \\
 \bottomrule
\end{tabular}
}
\end{table}

First, we investigated why there were different optimal aggregation types for the baseline methods and the proposed ones. According to Table~\ref{tab:units-per-word}, while the percentage of errors increases with the word length, their contribution to the model accuracy decreases faster. Therefore, aggregation types are of second importance after the per-frame sensitivity of a method. The maximum probability method is less sensitive than the Tsallis entropy-based, so its aggregation should be a product. $\alpha$-entropy-based methods can be tuned to rely less on aggregation.

Then we analyzed the ability of the chosen entropy-based methods to separate correct and incorrect word distributions. Figure~\ref{fig:hist-best-methods} contains histograms for metrics $\prod F^{e}_{g}(p)$, $\operatorname*{mean} F^{e}_{ts}(p)$, and $\min F^{e}_{ts}(p)$, measured on the LibriSpeech \textit{test-other} set. Larger total bar volume indicates more uniform confidence distribution. Different distribution shapes indicate better separability. All three methods cover the whole confidence spectrum. However, the entropy-based methods also transform word distributions themselves. The confidence metric $\prod F^{e}_{g}(p)$ made the incorrect word distribution uniform while correct words kept the trend towards 1.

For CTC, the confidence measure $\operatorname*{mean} F^{e}_{ts}(p)$ pushed correct words towards $1$, keeping incorrect words distributed uniformly. For RNN-T, the method pushed incorrect words towards $0$, holding the correct ones on a plateau from $0.4$ to $1$. The measure $\min F^{e}_{ts}(p)$ minimized confidence for incorrect words more intensively than the $mean$-aggregated method, but at the cost of a small correct word confidence leakage towards $0$. Also,  $\min F^{e}_{ts}(p)$ delivered the most similar numbers of words with edge confidence scores ($0$ for incorrect words and $1$ for correct ones) for both CTC and RNN-T.

Overall, exponentially normalized Tsallis entropy-based confidence with the $minimum$ aggregation was the best at separating correct and incorrect word distributions.

\begin{table}[!b]
\vspace{-10pt} 
\center
\small
\setlength{\tabcolsep}{2.5pt}
\caption{Different entropy-based confidence measures with $min$ aggregation for Conformer-RNN-T tested on LibriSpeech \textit{test-other}. Confidence measures: $F^{e}_{ts}(p)$ and $F^{e}_{r}(p)$ with $\alpha=1/3$, $F^{e}_{ts}(p)$ with $\alpha=1/4$ and $1/2$, $F_{ts}(p)$ and $F_{r}(p)$ with $\alpha=1/3$, and $F_{r}(p)$ with $\alpha=1/4$ and $1/2$. Metrics are in $\%$, except for $\mathrm{NCE}$.}
\vspace{-5pt} 
\label{tab:ent-measures}
\adjustbox{max width=\linewidth}{
\begin{tabular}{l|c|cccccccc}
    \toprule
    \multirow{2}{*}{\textbf{Method}}                    &
    \multirow{2}{*}{$\alpha$}                           &
    \multicolumn{3}{c}{$\mathrm{AUC}$}                  &
    $\mathrm{AUC}$ & $\mathrm{MAX}$ & $\mathrm{STD}$    &
    \multirow{2}{*}{$\mathrm{NCE}$}                     &
    \multirow{2}{*}{$\mathrm{ECE}$}                     \\
    \cmidrule(lr){3-5}
    \cmidrule(lr){6-8}
    & & $\mathrm{ROC}$ & $\mathrm{PR}$ & $\mathrm{NT}$ & \multicolumn{3}{c}{$\mathrm{YC}$} & &  \\
    \midrule
    $F^{e}_{ts}(p)$ & \multirow{2}{*}{$1/3$} & 85.85 & 98.90 & \textbf{47.17} & 30.71 & 62.07 & 18.42 & -2.22 & 28.95 \\
    $F^{e}_{r}(p)$ &  & 85.85 & 98.90 & \textbf{47.17} & 25.84 & 61.96 & 18.96 & -1.63 & 19.32 \\
    \cmidrule(lr){1-2}
    \multirow{2}{*}{$F^{e}_{ts}(p)$} & $1/4$ & \textbf{88.24} & \textbf{99.10} & 46.01 & \textbf{30.81} & \textbf{64.67} & \textbf{23.24} & -3.60 & 44.00 \\
     & $1/2$ & 81.72 & 98.60 & 43.92 & 26.29 & 51.48 & 14.35 & -1.46 & 26.77 \\
    \midrule
    $F_{ts}(p)$ & \multirow{2}{*}{$1/3$} & 85.85 & 98.90 & \textbf{47.17} & 00.87 & 44.13 & 04.90 & -0.16 & \textbf{02.91} \\
    $F_{r}(p)$ &  & 85.85 & 98.90 & \textbf{47.17} & 09.65 & 61.11 & 16.04 & \textbf{0.14} & 05.93 \\
    \cmidrule(lr){1-2}
    \multirow{2}{*}{$F_{r}(p)$} & $1/4$ & \textbf{88.24} & \textbf{99.10} & 46.01 & 12.98 & 64.54 & 19.23 & 0.09 & 06.94 \\
     & $1/2$ & 81.72 & 98.60 & 43.92 & 06.73 & 52.29 & 12.91 & 0.13 & 04.06 \\
 \bottomrule
\end{tabular}
}
\end{table}

Table~\ref{tab:ent-measures} contains a comparison of confidence  $\min F^{e}_{ts}(p)$ with $\alpha=1/3$ against other parametric measures and normalization approaches with the $min$ aggregation, measured for RNN-T on the \textit{test-other} set. Figure~\ref{fig:hist-ent-measures} serves to visualize the bias and separability of distributions for some of the methods. It can be seen that the R{\'e}nyi entropy-based confidence improves $\mathrm{NCE}$ and $\mathrm{ECE}$ for the cost of $\mathrm{AUC}_\mathrm{YC}$. The measure $F^{e}_{r}(p)$ with $\alpha=1/3$ assigns less zero confidence scores for incorrect words than it does ones for correct words.

$F^{e}_{ts}(p)$ with $\alpha=1/4$ is better than the baseline in most metrics except $\mathrm{NCE}$ and $\mathrm{ECE}$. This fact and Figure~\ref{fig:hist-rnnt-norm-ent-min-tsallis-exp-0.25} indicate that the scores of $F^{e}_{ts}(p)$ with $\alpha=1/4$ are underconfident (both distributions are pushed towards $0$). While underconfidence is a useful feature if confidence scores are used for classification, generally, low correct word scores are unnatural for an ASR system with high-accuracy. If $\alpha=1/4$ over-smoothes probabilities, then setting $\alpha=1/2$ is not enough for a method to obtain good classification capabilities and to separate correct and incorrect distributions well. Every considered confidence estimation variant with linear normalization had $\mathrm{AUC}_\mathrm{YC}$ score less than the $\mathrm{STD}_\mathrm{YC}$. This indicates that their confidence spectrum is narrower than $[0,1]$ (e.g. Figure~\ref{fig:hist-rnnt-norm-ent-min-renui-lin-0.25}), which reduces adjustability.

\vspace{-5pt} 
\section{Conclusion}
\label{sec:conclusion}
\vspace{-5pt} 

This paper presented a class of new fast non-trainable entropy-based confidence estimation methods for end-to-end ASR systems. We proposed new word-level confidence methods for CTC and RNN-T models using exponential normalization and mean and min aggregations of per-frame scores. Entropy-based confidence measures have similar computational complexity to confidence measures based on maximum per-frame probability, but they  better separate the confidence distributions of correct and incorrect words.

Our best confidence estimation method is based on the Tsallis entropy with $\alpha=1/3$, exponential normalization, and minimum aggregation function. It improved the incorrect words detection ability from $1.5$ to $4$ times over the maximum probability baseline for Conformer-CTC and Conformer-RNN-T. Moreover, the proposed confidence estimation method delivered similar metrics scores for CTC and RNN-T models. We analyzed entropy-based measures under various signal-to-noise ratios. We demonstrated that the estimator is noise robust and allows to filter out up to $40\%$ of model hallucinations on pure noise data at the cost of $5\%$ of correct words under regular acoustic conditions. Finally, we proposed new metrics based on the Youden's curve to better assess the adjustability of the methods.

\bibliographystyle{IEEEbib}
\bibliography{strings,refs}

\end{document}

%% file: nt_plot.tex
\begin{figure}[!b]
\vspace{-15pt} 
     \centering
     \begin{subfigure}[b]{0.8\linewidth}
         \centering
        \begin{tikzpicture}
        \begin{axis}[
            title=CTC,
            title style={at={(0.1,0.97)}},
            xlabel={SNR},
            ylabel={$\mathrm{AUC}_\mathrm{NT}$, $\%$},
            height=5cm,
            xmin=0, xmax=30,
            ymin=20, ymax=90,
            xtick={0,10,20,30},
            ytick={20,30,40,50,60,70,80,90},
            legend style={at={(0.0,0.0)},anchor=south west,nodes={scale=0.75, transform shape}},
            legend cell align=left,
            ymajorgrids=true,
            grid style=dashed,
        	enlargelimits=0.025,
        ]
        
        \addplot[
            color=red,
            mark=square,
            ]
            coordinates {
            (0,81.50)(10,63.02)(20,43.05)(30,35.17)};
            \addlegendentry{$\prod F_{max}(p)$}
        
        \addplot[
            color=brown,
            mark=halfcircle,
            ]
            coordinates {
            (0,83.36)(10,67.51)(20,48.75)(30,40.76)};
            \addlegendentry{$\prod F^{e}_{g}(p)$}
        
        \addplot[
            color=green,
            mark=triangle,
            ]
            coordinates {
            (0,78.11)(10,66.05)(20,55.63)(30,48.35)};
            \addlegendentry{$\operatorname*{mean} F^{e}_{ts}(p)$}
        
        \addplot[
            color=blue,
            mark=diamond,
            ]
            coordinates {
            (0,82.65)(10,70.78)(20,58.08)(30,50.29)};
            \addlegendentry{$\min F^{e}_{ts}(p)$}
        
        \end{axis}
        \end{tikzpicture}
     \end{subfigure}
     \begin{subfigure}[b]{0.8\linewidth}
         \centering
        \begin{tikzpicture}
        \begin{axis}[
            title=RNNT,
            title style={at={(0.1,0.97)}},
            xlabel={SNR},
            ylabel={$\mathrm{AUC}_\mathrm{NT}$, $\%$},
            height=5cm,
            xmin=0, xmax=30,
            ymin=20, ymax=90,
            xtick={0,10,20,30},
            ytick={20,30,40,50,60,70,80,90},
            legend style={at={(1.0,1.0)},anchor=north east,nodes={scale=0.75, transform shape}},
            legend cell align=left,
            ymajorgrids=true,
            grid style=dashed,
        	enlargelimits=0.025,
        ]
        
        \addplot[
            color=red,
            mark=square,
            ]
            coordinates {
            (0,62.33)(10,40.99)(20,27.18)(30,22.34)};
            \addlegendentry{$\prod F_{max}(p)$}
        
        \addplot[
            color=brown,
            mark=halfcircle,
            ]
            coordinates {
            (0,68.92)(10,48.33)(20,33.64)(30,28.93)};
            \addlegendentry{$\prod F^{e}_{g}(p)$}
        
        \addplot[
            color=green,
            mark=triangle,
            ]
            coordinates {
            (0,70.49)(10,55.28)(20,44.55)(30,36.55)};
            \addlegendentry{$\operatorname*{mean} F^{e}_{ts}(p)$}
        
        \addplot[
            color=blue,
            mark=diamond,
            ]
            coordinates {
            (0,77.63)(10,65.56)(20,56.20)(30,49.91)};
            \addlegendentry{$\min F^{e}_{ts}(p)$}
        
        \end{axis}
        \end{tikzpicture}
     \end{subfigure}
\vspace{-10pt} 
\caption{Evaluation of confidence methods: $\prod F_{max}(p)$, $\prod F^{e}_{g}(p)$, $\operatorname*{mean} F^{e}_{ts}(p)$, and $\min F^{e}_{ts}(p)$, under different signal-to-noise -ratio. Metric: $\mathrm{AUC}_\mathrm{NT}$. Conformer-CTC and Conformer-RNNT, LS \textit{test-other}.}
\label{fig:nt-plot}
\end{figure}

%% file: main.bbl
\begin{thebibliography}{10}

\bibitem{vesely2013asru}
K.~Vesel{\`y}, M.~Hannemann, and L.~Burget,
\newblock ``Semi-supervised training of deep neural networks,''
\newblock in {\em ASRU}, 2013, pp. 267--272.

\bibitem{riccardi2005}
G.~Riccardi and D.~Hakkani-Tur,
\newblock ``Active learning: theory and applications to automatic speech
  recognition,''
\newblock {\em IEEE Transactions on Speech and Audio Processing}, vol. 13, no.
  4, pp. 504--511, 2005.

\bibitem{yu2010csl}
D.~Yu, B.~Varadarajan, L.~Deng, and A.~Acero,
\newblock ``Active learning and semi-supervised learning for speech
  recognition: A unified framework using the global entropy reduction
  maximization criterion,''
\newblock {\em Computer Speech \& Language}, vol. 24, no. 3, pp. 433--444,
  2010.

\bibitem{drugman2016}
T.~Drugman, J.~Pylkkönen, and R.~Kneser,
\newblock ``Active and semi-supervised learning in {ASR}: Benefits on the
  acoustic and language models,''
\newblock in {\em INTERSPEECH}, 2016.

\bibitem{uebel2001speaker}
L.~F. Uebel and P.~C. Woodland,
\newblock ``Speaker adaptation using lattice-based {MLLR},''
\newblock in {\em ITRW}, 2001.

\bibitem{zbib2019aS}
R.~Zbib, L.~Zhao, D.~Karakos, W.~Hartmann, J.~DeYoung, Z.~Huang, Z.~Jiang,
  N.~Rivkin, L.~Zhang, R.~Schwartz, and J.~Makhoul,
\newblock ``Neural-network lexical translation for cross-lingual {IR} from text
  and speech,''
\newblock in {\em SIGIR}, 2019, pp. 645--654.

\bibitem{saleem2004using}
S.~Saleem, S.-C. Jou, S.~Vogel, and T.~Schultz,
\newblock ``Using word lattice information for a tighter coupling in speech
  translation systems,''
\newblock in {\em ICSLP}, 2004.

\bibitem{besacier-etal-2014-word}
L.~Besacier, B.~Lecouteux, N.~Q. Luong, K.~Hour, and M.~Hadjsalah,
\newblock ``Word confidence estimation for speech translation,''
\newblock in {\em International Workshop on Spoken Language Translation}, 2014.

\bibitem{wessel2001}
F.~Wessel, R.~Schluter, K.~Macherey, and H.~Ney,
\newblock ``Confidence measures for large vocabulary continuous speech
  recognition,''
\newblock {\em IEEE Transactions on Speech and Audio Processing}, vol. 9, no.
  3, pp. 288--298, 2001.

\bibitem{Jiang2005ConfidenceMF}
H.~Jiang,
\newblock ``Confidence measures for speech recognition: A survey,''
\newblock {\em Speech Communication}, vol. 45, pp. 455--470, 2005.

\bibitem{Yu2011CalibrationOC}
D.~Yu, J.~Li, and L.~Deng,
\newblock ``Calibration of confidence measures in speech recognition,''
\newblock {\em IEEE Transactions on Audio, Speech, and Language Processing},
  vol. 19, pp. 2461--2473, 2011.

\bibitem{graves_connectionist_2006}
A.~Graves, S.~Fernández, F.~Gomez, and J.~Schmidhuber,
\newblock ``Connectionist temporal classification: labelling unsegmented
  sequence data with recurrent neural networks,''
\newblock in {\em ICML}, 2006.

\bibitem{graves2012transducer}
A.~Graves,
\newblock ``Sequence transduction with recurrent neural networks,''
\newblock in {\em ICML}. 2012, Workshop on Representation Learning.

\bibitem{hendrycks2016iclr}
D.~Hendrycks and K.~Gimpel,
\newblock ``A baseline for detecting misclassified and out-of-distribution
  examples in neural networks,''
\newblock in {\em ICLR}, 2016.

\bibitem{park20d_interspeech}
D.~S. Park, Y.~Zhang, Y.~Jia, W.~Han, C.-C. Chiu, B.~Li, Y.~Wu, and Q.~V. Le,
\newblock ``{Improved Noisy Student Training for Automatic Speech
  Recognition},''
\newblock in {\em INTERSPEECH}, 2020.

\bibitem{nguyen2015}
A.~Nguyen, J.~Yosinski, and J.~Clune,
\newblock ``Deep neural networks are easily fooled: High confidence predictions
  for unrecognizable images,''
\newblock in {\em CVPR}, 2015.

\bibitem{vyas2019icassp}
A.~Vyas, P.~Dighe, S.~Tong, and H.~Bourlard,
\newblock ``Analyzing uncertainties in speech recognition using dropout,''
\newblock in {\em ICASSP}, 2019.

\bibitem{malinin2021uncertainty}
A.~Malinin and M.~Gales,
\newblock ``Uncertainty estimation in autoregressive structured prediction,''
\newblock in {\em ICLR}, 2021.

\bibitem{Oneata2021}
D.~Oneaţă, A.~Caranica, A.~Stan, and H.~Cucu,
\newblock ``An evaluation of word-level confidence estimation for end-to-end
  automatic speech recognition,''
\newblock in {\em SLT}, 2021.

\bibitem{li2020confidence}
Q.~Li, D.~Qiu, Y.~Zhang, B.~Li, Y.~He, P.~C. Woodland, L.~Cao, and T.~Strohman,
\newblock ``Confidence estimation for attention-based sequence-to-sequence
  models for speech recognition,''
\newblock in {\em ICASSP}, 2021.

\bibitem{jeon2020}
W.~Jeon, M.~Jordan, and M.~Krishnamoorthy,
\newblock ``On modeling {ASR} word confidence,''
\newblock in {\em ICASSP}, 2020.

\bibitem{woodward2020confidence}
A.~Woodward, C.~Bonn{\'\i}n, I.~Masuda, D.~Varas, E.~Bou-Balust, and
  J.~Riveiro,
\newblock ``Confidence measures in encoder-decoder models for speech
  recognition,''
\newblock in {\em INTERSPEECH}, 2020.

\bibitem{qiu2021learning}
D.~Qiu, Q.~Li, Y.~He, Y.~Zhang, B.~Li, L.~Cao, R.~Prabhavalkar, D.~Bhatia,
  W.~Li, K.~Hu, T.N. Sainath, and I.~McGraw,
\newblock ``Learning word-level confidence for subword end-to-end {ASR},''
\newblock in {\em ICASSP}, 2021.

\bibitem{li2021residual}
Q.~Li, Y.~Zhang, B.~Li, L.~Cao, and P.~Woodland,
\newblock ``Residual energy-based models for end-to-end speech recognition,''
\newblock in {\em INTERSPEECH}. Aug 2021, ISCA.

\bibitem{qui2021multi}
D.~Qiu, Y.~He, Q.~Li, Y.~Zhang, L.~Cao, and I.~McGraw,
\newblock ``Multi-task learning for end-to-end {ASR} word and utterance
  confidence with deletion prediction,''
\newblock in {\em INTERSPEECH}, 2021.

\bibitem{wang2021word}
M.~Wang, H.~Soltau, L.~Shafey, and I.~Shafran,
\newblock ``Word-level confidence estimation for {RNN} transducers,''
\newblock in {\em ASRU}, 2021.

\bibitem{evermann2000posterior}
G.~Evermann and P.C. Woodland,
\newblock ``Posterior probability decoding, confidence estimation and system
  combination,''
\newblock in {\em {NIST} Speech Transcription Workshop}. Citeseer, 2000,
  vol.~27, pp. 78--81.

\bibitem{zapotoczny2019}
M.~Zapotoczny, P.~Pietrzak, A.~Łańcucki, and J.~Chorowski,
\newblock ``{Lattice Generation in Attention-Based Speech Recognition
  Models},''
\newblock in {\em INTERSPEECH}, 2019, pp. 2225--2229.

\bibitem{kuchaiev2019nemo}
O.~Kuchaiev, J.~Li, H.~Nguyen, O.~Hrinchuk, R.~Leary, B.~Ginsburg, S.~Kriman,
  S.~Beliaev, V.~Lavrukhin, J.~Cook, et~al.,
\newblock ``{NeMo}: a toolkit for building {AI} applications using neural
  modules,''
\newblock {\em arXiv:1909.09577}, 2019.

\bibitem{tornetta_2021}
G.~N. Tornetta,
\newblock ``Entropy methods for the confidence assessment of probabilistic
  classification models,''
\newblock {\em Statistica}, vol. 81, no. 4, pp. 383–398, 2021.

\bibitem{phan2022sleeptransformer}
H.~Phan, K.B. Mikkelsen, O.~Chen, P.~Koch, A.~Mertins, and M.~De~Vos,
\newblock ``Sleeptransformer: Automatic sleep staging with interpretability and
  uncertainty quantification,''
\newblock {\em IEEE Transactions on Biomedical Engineering}, 2022.

\bibitem{tsallis1988possible}
C.~Tsallis,
\newblock ``Possible generalization of {Boltzmann-Gibbs} statistics,''
\newblock {\em Journal of statistical physics}, vol. 52, no. 1, pp. 479--487,
  1988.

\bibitem{hinton2015}
G.~Hinton, O.~Vinyals, and J.~Dean,
\newblock ``Distilling the knowledge in a neural network,''
\newblock in {\em NIPS Deep Learning and Representation Learning Workshop},
  2015.

\bibitem{renyi1961measures}
A.~R{\'e}nyi et~al.,
\newblock ``On measures of entropy and information,''
\newblock in {\em Proceedings of the fourth Berkeley symposium on mathematical
  statistics and probability}. Berkeley, California, USA, 1961, vol.~1, pp.
  547--561.

\bibitem{Gulati2020}
A.~Gulati, J.~Qin, C.-C. Chiu, N.Parmar, Y.~Zhang, J.~Yu, W.~Han, S.~Wang,
  Z.~Zhang, Y.~Wu, and R.~Pang,
\newblock ``{Conformer: Convolution-augmented Transformer for Speech
  Recognition},''
\newblock in {\em INTERSPEECH}, 2020.

\bibitem{panayotov2015librispeech}
V.~Panayotov, G.~Chen, D.~Povey, and S.~Khudanpur,
\newblock ``{LibriSpeech}: an {ASR} corpus based on public domain audio
  books,''
\newblock in {\em ICASSP}, 2015.

\bibitem{font2013freesound}
F.~Font, G.~Roma, and X.~Serra,
\newblock ``{Freesound} technical demo,''
\newblock in {\em Proceedings of the 21st ACM international conference on
  Multimedia}, 2013.

\bibitem{snyder2015musan}
D.~Snyder, G.~Chen, and D.~Povey,
\newblock ``Musan: A music, speech, and noise corpus,''
\newblock {\em arXiv:1510.08484}, 2015.

\bibitem{siu97_eurospeech}
M.~Siu, H.~Gish, and F.~Richardson,
\newblock ``{Improved estimation, evaluation and applications of confidence
  measures for speech recognition},''
\newblock in {\em Eurospeech}, 1997, pp. 831--834.

\bibitem{naeini2015obtaining}
M.~Naeini, G.~Cooper, and M.~Hauskrecht,
\newblock ``Obtaining well calibrated probabilities using bayesian binning,''
\newblock in {\em AAAI Conference on Artificial Intelligence}, 2015.

\bibitem{youden1950}
W.~J. Youden,
\newblock ``Index for rating diagnostic tests,''
\newblock {\em Cancer}, vol. 3, no. 1, pp. 32--35, 1950.

\end{thebibliography}
